\documentclass[twocolumn]{pasj00}

\title{Suzaku Results on Cygnus X-1 in the Low/Hard State}
\author{Kazuo \textsc{Makishima},\altaffilmark{1,2}
Hiromitsu \textsc{Takahashi},\altaffilmark{3} 
Shin'ya \textsc{Yamada},\altaffilmark{1}
Chris \textsc{Done,}\altaffilmark{4}\\
Aya \textsc{Kubota,}\altaffilmark{5}
Tadayasu \textsc{Dotani},\altaffilmark{6}
Ken \textsc{Ebisawa},\altaffilmark{6}
Takeshi \textsc{Itoh},\altaffilmark{1}
Shunji \textsc{Kitamoto},\altaffilmark{7}\\
Hitoshi \textsc{Negoro},\altaffilmark{8}
Yoshihiro \textsc{Ueda},\altaffilmark{9}
and Kazutaka \textsc{Yamaoka},\altaffilmark{10}\\
}
\altaffiltext{1}{
   Department of Physics, The University of Tokyo,
   7-3-1, Hongo, Bunkyo-ku, Tokyo, 113-0033, Japan}
\altaffiltext{2}{
   Cosmic Radiation Laboratory, Institute of Physical and Chemical 
   Research (RIKEN)\\
   2-1 Hirosawa, Wako-shi, Saitama, 351-0198, Japan}
\altaffiltext{3}{
Department of Physical Science, School of Science, Hiroshima University\\
1-3-1 Kagamiyama, Higashi-Hiroshima, Hiroshima 739-8526, Japan}
\altaffiltext{4}{
Durham University, United Kingdom}
\altaffiltext{5}{
Department of Electronic Information Systems, Shibaura Institute of Technology\\
307 Fukasaku, Minuma-ku, Saitama-shi, Saitama, 337-8570, Japan}
\altaffiltext{6}{
   Institute of Space and Astronautical Science, JAXA, \\
   3--1--1, Yoshinodai, Sagamihara, Kanagawa,  229-8510, Japan}
\altaffiltext{7}{
   Department of Physics, Rikkyo University, \\
   3-34-1, Nishi-Ikebukuro, Toshima--ku, Tokyo 171--8501, Japan}
\altaffiltext{8}{
   Department of Physics, College of Science and Technology, Nihon
   University\\
   1-8, Kanda-Surugadai, Chiyoda-ku, Tokyo, 101-8308, Japan}
\altaffiltext{9}{
   Department of Astronomy, Kyoto University\\
    Kitashirakawa-Oiwake-cho, Sakyo-ku, Kyoto, 606-8502, Japan}
\altaffiltext{10}{
   Department of Physics \& Mathematics, Aoyama Gakuin University\\
   5-10-1 Sagamihara, Kanagawa, 229-8558, Japan}
\email{maxima@phys.s.u-tokyo.ac.jp}

\Received{$\langle$reception date$\rangle$}
\Accepted{$\langle$acception date$\rangle$}
\Published{$\langle$publication date$\rangle$}
\begin{document}
\maketitle

\begin{abstract}
The black-hole binary Cygnus X-1 was observed for 17 ks 
with the Suzaku X-ray observatory in 2005 October, 
while it was in a low/hard state
with a  0.7--300  keV luminosity  of $4.6\times 10^{37}$ erg s$^{-1}$.
The  XIS and HXD spectra, spanning 0.7--400 keV,
were reproduced successfully incorporating a cool accretion disk and a hot Comptonizing corona.
The  corona is characterized by an electron temperature of  $ \sim 100$ keV,
and two optical depths of $\sim 0.4$ and $\sim 1.5$
which account for the harder and softer continua, respectively.
The disk has the innermost temperature of $\sim 0.2$ keV,
and is though  to protrude  half way into the corona.
The disk not only provides seed photons to the Compton cloud,
but also produces a soft spectral excess, 
a mild reflection hump, and a weakly broadened iron line.
A comparison with the Suzaku data on GRO~J1655$-$40 reveals
several interesting spectral differences,
which can mostly be attributed to inclination effects
assuming that the disk has a flat geometry while the corona is grossly spherical.
An intensity-sorted spectroscopy indicates
that the continuum becomes less Comptonized 
when the source flares up on times scales of 1--200 s,
while the underlying disk  remains unchanged.
\end{abstract}


\section{Introduction}
\label{sec:intro}

Luminous soft X-ray radiation of accreting stellar-mass black holes (BHs)
has generally been explained as  thermal emission  
from optically-thick (in particular ``standard'') accretion disks
(Shakura \& Sunyaev 1973; 
\cite{Makishima1986,Dotani1997,RM06}),
which are expected to form around them
under rather high accretion rates.
In contrast, their hard X-ray production process is much less understood,
even though intense hard X-ray emission  characterizes black-hoe binaries (BHBs)
among various types of  compact X-ray sources  in the Milky Way and Magellanic clouds.

Indeed,  BHBs  often emit a major fraction 
of their radiative luminosity in the hard X-ray band,
in the form of spectral hard-tail  component
if  they are in so-called high/soft state,
or  as the entire power-law (hereafter PL) like continua 
if they are in so-called low/hard state (hereafter LHS)
which appears under relatively low accretion rates.
Furthermore, the hard X-ray emission 
(particularly in the LHS) involves another interesting aspect,
namely the long-known aperiodic variation over a wide frequency range 
(e.g., \cite{Oda1971,Oda1977,Noloan1981,Miyamoto1991,Pottschmidt2003,RM06}).
These spectral and timing studies are not limited to stellar-mass BHs,
since a fair fraction of active galactic nuclei (AGNs) are also 
considered to be in a state analogous to the LHS of BHBs,
emitting luminous hard X-rays (e.g., \cite{MDK1993}).

In hard X-rays,
a  BHB in the LHS emits a roughly PL shaped continuum
with a photon index $\Gamma \sim 1.7$ (e.g., \cite{RM06}),
but the spectrum gradually steepens toward higher energies.
This has been explained  in terms of
unsaturated Comptonization of some soft seed photons,
in a cloud of hot Maxwellian electrons, or a ``corona",
with a  temperature of $T_{\rm e}=50-100$ keV
and an optical depth of $\tau \sim 1$ 
\citep{SLE1976,ST1979,ST1980,G1997,DiSalvo2001,Frontera2001,ZG2004,Ibragimov2005}.
As a result, the Comtpon $y$-parameter,
$y \equiv 4\tau kT_{\rm e}/m_{\rm e} c^2$,
also take a value of $\sim 1$,
where $k$ is the Boltzmann constant,
$m_{\rm e}$ is the electron mass and $c$ is the light velocity.
The electrons will be  cooled 
by transferring  their energies to the  photons,
while heated via Coulombic collisions by ions
that acquire nearly a free-fall temperature (e.g., \cite{ZLS1999}).
The Comptonizing cloud may be identified with 
radiatively inefficient inner accretion flow 
(e.g., \cite{Ichimaru1977, ADAF95}).

Superposed on the dominant continuum characterizing the LHS,
we often observe a spectral hump at energies of  20--40 keV
due to  ``reflection'' of the primary X-rays  
by some cool materials  (e.g., \cite{LW1988,Inoue1989}),
a fluorescent Fe-K emission line (\cite{Basko1978,Kitamoto1990}),
and an apparently ``smeared'' Fe-K edge (\cite{Ebisawa1994}).
These features can be explained by assuming
that a fraction of the  primary hard X-rays are reprocessed
by  cool materials located near the BH,
presumably an optically-thick part of the  accretion disk  
(\cite{Done1992,Marshall1993,Ebisawa1996,G1997}).
This part of the disk may  at the same time produce an X-ray excess 
which we often observe at the softest end of the spectrum (e.g., \cite{BH1991}),
and furthermore,
may provide the Comptonizing corona with the seed photons  (e.g., \cite{SLE1976}).

In spite of general success of this currently popular interpretation
of the LHS in terms of Comptonizing hot clouds and cool materials,
details of the scenario still need to be elucidated.
It is  yet to be confirmed
that the seed  photons are provided by a cool  accretion disk
(e.g., \cite{ZLS1999}),
rather than by thermal cyclotron emission in the Compton corona itself (e.g., \cite{Kan2001}).
The geometry and size of the Comptonizing cloud is not well known,
except for loose constraints, including;
that the cloud should be large enough for the compactness parameter
to remain within a reasonable range (\cite{LZ1987}),
but should not be too large to be able to produce
the hard X-ray variation on a $\sim 1$ s time scale,
and that it should have a sufficient vertical height
so as to efficiently illuminate the reprocessing matter.
Still unsettled is the relation between  the hot corona and the putative cool disk;
whether the cool disk is  truncated at a large radius \citep{Z1998, DZ1999,Z2004},
or it penetrates deep into the hot corona \citep{Young2001,Miller2006}.
Yet another  fundamental issue is
how to understand  the violent hard X-ray variability
in the context of thermal Comptonization scenario (e.g., \cite{Miyamoto1987}).

To address these issues associated with the LHS of accreting BHs,
a number of spectral studies have been conducted so far,
by fitting observed broad-band spectra 
with theoretical spectral models (e.g., \cite{G1997,Z1998,Dove1998,Frontera2001}).
Similarly, extensive studies in the frequency domain
have been conducted to characterize the random variations
(e.g., \cite{MK1989,NKM1994,Nowak1999b,Gilfanov2000,Revnivtsev2000,RM06}).
However, the former approach is limited 
in that different modelings often degenerate.
The latter approach has also a limitation,
in that it may not  constrain
details of the Comptonization scenario
without invoking particular models \citep{KHT1997,HKT1997}.

We analyze broad-band (0.7--400 keV) Suzaku data of Cygnus X-1 
(Cyg X-1; Oda 1977; Lian and Nolan 1984) acquired in 2005 October.
With the wide-band sensitivity,
Suzaku is expected to provide  novel diagnostics of this  BHB,
the first celestial object  suspected  to be a black hole  \citep{Oda1971}.
To overcome the above difficulties,
we  compare the  Suzaku spectra of Cyg X-1 with those of  GRO~1655$-$40,
searching the two spectra for differences attributable
to their different system inclination,
 $i \sim 45^\circ$ of Cyg X-1 (\cite{A2004})
and  $i \sim 70^\circ$  of GRO~1655$-$40  (\cite{G2001}).
We hence refer  to the Suzaku results on  
GRO~1655$-$40 by \citet{Takahashi1655} as Paper I.
We also analyze spectral changes associated with the random 
variations on a time scale of $\sim 1$ s (section~\ref{sec:ana_varspec}),
to see how the Compton parameters are changing;
this will be  complementary to the study of long-term spectral changes (e.g.,\cite{Ibragimov2005}).
We employ  a distance of  2.5 kpc \citep{Margon1973,Bregman1973} to Cyg X-1.
The errors refer to statistical 90\% confidence limits.

\section{Observation and Data Processing}
\label{sec:obs}

\subsection{Observation}
\label{subsec:obs}

The present Suzaku observation of Cyg X-1 was performed  on 2005 October 5, 
from UT 04:34  through 15:11 for  a net exposure of 17.4~ks, 
as a part of the initial performance verification program of Suzaku \citep{Mitsuda2007}.
Suzaku carries  4 sets of X-ray telescopes \citep{XRT2007},
each coupled to a focal-plane X-ray CCD camera called  XIS
(X-ray Imaging Spectrometer; \cite{KoyamaXIS}) 
operating in the energy range of 0.2--12~keV.
Three  (XIS0, XIS2, and XIS3) of the four XIS sensors use front-illuminated (FI) CCDs, 
while XIS1 utilizes  a back-illuminated (BI) one
thus achieving an improved soft X-ray response. 
The Hard X-ray Detector (HXD; \cite{TakahashiHXD, KokubunHXD})
covers the 10--70 keV energy band with Si PIN photo-diodes  (HXD-PIN), 
and the 50--600 keV range with GSO scintillation counters (HXD-GSO).

Because the optical axes of the XIS and the HXD are slightly ($\sim 3'.5$) offset,
the observation was carried out with the source  placed 
at the center of the XIS field of view. 
The HXD was operated in the nominal mode throughout the observation. 
Because the source was very bright, 
the XIS was operated with  the ``1/8 window option'', 
in which a smaller field of view of $17^\prime.8\times 2^\prime.2$ is read out every 1~s.  
The editing mode of the FI CCDs was set to $2\times2$,
while that of the BI CCD (XIS1) was  $3\times3$.
In the present observation, however,  
XIS1 with the BI CCD chip suffered severe telemetry saturation,
for more than 60\% of the total exposure. 

During the observation, 
the averaged source count rates (excluding backgrounds) 
of XIS0, HXD-PIN, and HXD-GSO were
$\sim 300$,  $\sim 47$, and $\sim 20$ c~s$^{-1}$,
in the energy bands of 0.5--10~keV, 10--70~keV, and 70--400~keV, respectively.
However, the true XIS0 counts would be higher,
if the data  were free from  telemetry overflows and event pile up
(subsection~\ref{subsec:xisreduction}).

\subsection{XIS Data Selection
\label{subsec:xisreduction}}

We retrieved  the XIS data which were prepared via  version 1.2 processing.
The XIS data were further screened based on the following standard criteria: 
the XIS GRADE should be 0, 2, 3, 4, or  6;
the time interval after an exit from the South Atlantic Anomaly should be longer than 500 s;
and the object should be at least $5^\circ$ and $20^\circ$
above the dark and sunlit Earth rim, respectively.

In order to conduct spectral and timing studies,
we utilize  XIS events extracted from a region
which is defined to fit in the ``1/8 window''.
Since the Suzaku attitude is known to wobble on a time scale of
several tens minutes, 
the region needs to be a little smaller than the window size.
Considering the magnitude of the wobbling, 
we defined the region as an intersection 
between a rectangle of $70\times1024$ pixels
and a circle with a radius of $3'.3$ (200 pixels),
both centered always on the image peak.
For this purpose, we calculated the image center every 200~s
and re-defined the event extraction region also every 200 s.
Thus, the location of the region on the CCD chip is time dependent.
The background was entirely negligible throughout the observation
(less than 0.1\% of the signal at 8~keV considering the Galactic ridge emission),
so was not subtracted from the data.

Because Cyg X-1 was so bright that we need to carefully remove
the effects of telemetry saturation and photon pile up.
When the number of detected events exceeds 
the telemetry allocation to the XIS, 
the excess events are discarded without being transferred to telemetry.
This phenomenon is called telemetry saturation.
Since the event buffer is cleared every 8~s,
and the XIS exposure was 1~s in the present observation,
we excluded the telemetry-saturated data in the following way.
For each 8-s window, 
we calculated an average count rate over the first 4 s
which are usually free from the telemetry saturation.
If a subsequent exposure in the same 8-s window showed a 
count rate lower than this average by  more than 2 standard deviations, 
we judged that the telemetry was saturated, 
and discarded this and the subsequent exposures in the same window.
This typically occurred in the last one or  two exposures in the 8-s window.
Because these telemetry-saturated time bins are regarded as dead times,
the source count rate remains approximately  intact,
although  the exposure is reduced.

After filtering out the telemetry-saturated data, 
we further removed the effects of photon pile up.
Although the Suzaku X-ray telescopes have relatively wide point spread functions,
photon pile up becomes noticeable at the image center
when observing a point source brighter than $\sim 100$ cts per exposure.
The average count rate of Cyg X-1 during the present observation
was $\sim\!3$ times higher  than this limit.
We therefore accumulated XIS events by excluding
circular regions of various radii at the image center,
and examined the obtained series of spectra for shape changes.
As a result, we have decided to exclude an image-center
region with a radius of $1'.0$ (60 pixels).
Although this means that the event extraction region has 
a complicated shape (intersection of an annulus and a rectangle), 
we use standard ancillary response files
(\texttt{ae\_xi0\_xisnom4\_20060615.arf}, etc.)
defined for a circular region of  $4'.0$ radius.
This is because the point spread function of the telescope has little energy dependence,
and hence the spectrum would not change except in normalization.
We confirmed this inference using the data of a moderately bright point source, 
SS Cyg observed on 2005 November 2,
and found that the spectral normalization should be multiplied
by a factor of $\sim 0.092$ compared with that of  PIN,
when excluding the central $1'$ of the image;
this is taken into account in the subsequent spectral analysis.
We also use the standard response matrices 
(\texttt{ae\_xi0\_20060213.rmf}, etc.) for the XIS.

\subsection{HXD Data Selection and Background Subtraction
\label{subsec:hxdreduction}}

The HXD data utilized here were  prepared via version 1.2 processing.
We further screened the HXD events by imposing conditions 
that more than 500 s should be elapsed  
from an exit out of  the South Atlantic Anomaly,
that the target should be $> 5^\circ$ above the Earth rim, 
and that the geomagnetic cutoff rigidity should be higher than 8 GV. 

Since the HXD has neither imaging capability nor an offset counter
to measure instantaneous non X-ray background (NXB),
we need to estimate the NXB and subtract it from the raw data.
Accordingly, we employed the PIN background modeling by \citet{WatanabePINbgd},
known as ``PINUDLC (bgd\_a)'' model
\footnote{http://www.astro.isas.jaxa.jp/suzaku/analysis/hxd/hxdnxb/},
which combines actual HXD-PIN data acquired 
while pointing to night Earth \citep{KokubunHXD}.
Thus, we produced a large number (10.0 times as large as the actual)
of fake non X-ray background events,
which would just be detected during the present observation
if we had an imaginary NXB monitor detector.
By processing this fake event files exactly in the same manner as the actual data,
we can estimate NXB contributions in the PIN data,
and subtract them from PIN light curves as well as from PIN spectra,
with a typical accuracy of 5\% or better.
The cosmic X-ray background entering through the PIN field of view
is entirely negligible in the present study.

Subtraction of the HXD-GSO background employs a different method,
developed by \citet{FukazawaGSObgd} and called  ``LCFIT (bgd\_d)'' model
\footnote{http://www.astro.isas.jaxa.jp/suzaku/analysis/hxd/gsonxb/}.
In this case, the whole GSO energy band is divided into 32 energy bins,
and a light curve of each energy bin is produced over a long period.
Because each light curve has characteristic repetition patterns 
with periods of one orbital revolution,
one day (due to the relation with respect to the South Atlantic Anomaly),
and 51 days (due to orbital precession),
it can be expressed as a function of time 
using a relatively small number of adjustable parameters.
The parameters can be determined by analyzing these light curves
over a long period, typically one month.
We can thus  derive the expected NXB counts in individual energy bins
at a given time during the present observation,
and subtract them either from light curves or spectra.
The GSO data are hence analyzed using the same energy binnings 
as used in the NXB modeling.
When integrated over $\sim 1$ day,
this method reproduces the GSO background  
within a systematic error of $\sim 2\%$ \citep{TakahashiGSObgd}. 
Since the PIN and GSO background models 
both utilize PIN upper-discriminator hit rates,
their basic time resolution is typically 4 s or longer,
depending on the telemetry rate.

\bigskip
\begin{figure}[hbt]
\centerline{
\FigureFile(9cm,){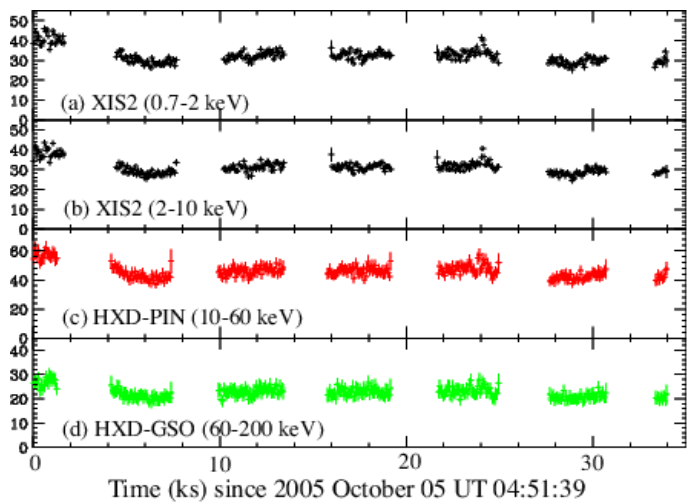}
}
\caption{Light curves of Cyg X-1 binned into 64 s,
obtained in the present Suzaku observation performed in 2005 October.
The ordinates are in units of ct s$^{-1}$.
From top to bottom, panels refer to  those obtained 
with XIS2 (0.7--2 keV), also with XIS2 (2--10 keV),
with HXD-PIN (10--60 keV), and with HXD-GSO (60--200 keV).
The HXD backgrounds were subtracted.
The periodic data gaps are due to  spacecraft passages
through the South Atlantic Anomaly and the Earth occultation.
}
\label{fig:ltcurves}
\end{figure}

\subsection{Light Curves
\label{subsec:obs_ltcurve}}

Figure~\ref{fig:ltcurves} shows XIS2, HXD-PIN, and HXD-GSO
light curves of Cyg X-1 from the present observation.
The HXD backgrounds (PIN and GSO) were subtracted
in the way as described in section~\ref{subsec:hxdreduction},
and the results were corrected for dead times.
In contrast,  the XIS background, which is completely negligible, is inclusive.
The count rates indicated by these XIS light curves are 
by an order of mangnitude lower than those actually observed,
because of the image-center exclusion
which was needed to avoid pile-up events
(subsection \ref{subsec:xisreduction}).
On time scales of a few minutes  to a day,
the source thus varied randomly typically by $\sim \pm 10$\%.

\subsection{Time-Averaged Spectra
\label{subsec:obs_avspec}}

\begin{figure}[hbt]
\centerline{
\FigureFile(8cm,){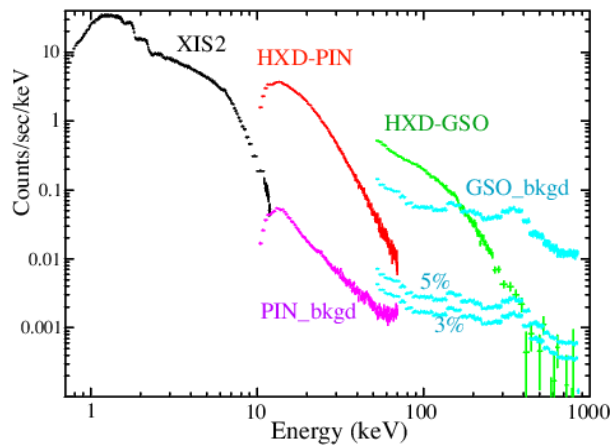}
}
\caption{
Time-averaged  XIS2 (black),  HXD-PIN (red), 
and HXD-GSO (green) spectra of Cyg X-1,
shown after  subtracting the background 
but not removing the instrumental responses.
The XIS2 spectrum is accumulated after removing the telemetry overflow 
and and cutting off the image center (see text).
The modeled non X-ray background of HXD-PIN is indicated in magenta.
That of HXD-GSO is shown in cyan,
together with its 5\% and 3\% levels.}
\label{fig:avspec}
\end{figure}

We show in figure~\ref{fig:avspec}
the time-averaged Suzaku spectra of Cyg X-1.
Like in section~\ref{subsec:obs_ltcurve},
we corrected the HXD spectra  for the dead times,
and subtracted the  backgrounds using the method 
described in section~\ref{subsec:hxdreduction}.
In contrast, the XIS2 spectrum is here and hereafter
not corrected for the normalization reduction
(down to 0.092 of the original level),
due to  the exclusion of the central $1'$ of the images
(subsection \ref{subsec:xisreduction}).
For comparison,
we superpose the modeled background spectra
of HXD-PIN and HXD-GSO.

As seen  in figure~\ref{fig:avspec},
we have significantly detected Cyg X-1 
nearly across the full energy band of Suzaku.
In particular, the source signals in HXD-PIN 
exceed the NXB by more than a factor of $\sim 5$,
even at the highest energy of $\sim 70$ keV.
Therefore, the systematic background uncertainty of HXD-PIN
is completely negligible  compared to photon counting statistics.
Even in the GSO range,
the source signal is so strong 
that it can be detectable up to $\sim 400$ keV,
above which the systematic background errors start dominating.

\section{Analysis of the Time Averaged Spectra}
\label{sec:ana_avspec}
This section is devoted to the study of 
time-averaged Suzaku spectra of Cyg X-1.
Because of the strong telemetry saturation (section \ref{subsec:obs}),
we do not use the  data from XIS1 (BI).
Of the three FI CCD cameras comprising the XIS,
only the data from XIS2 are employed in our spectral analyses,
while those from XIS0 and XIS3 are utilized 
in section~\ref{sec:ana_varspec} for a different purpose;
even if  we incorporated data from the other XIS sensors,
systematic errors would dominate
before data statistics are improved,
because Cyg X-1 is so bright for the XIS.

\subsection{Fitting with conventional models}
\label{subsec:convfits}

In order to roughly characterize the time-averaged 
broad-band spectra derived in subsection~\ref{subsec:obs_avspec},
we simultaneously fitted simple empirical models 
to the XIS2, HXD-PIN, and HXD-GSO spectra,
so that the results may be compared with other observations of Cyg X-1,
as well as those of other similar objects.
Here and hereafter, 
the overall model normalization is allowed to differ between the XIS and HXD
(mainly to absorb uncertainties introduced by 
the removal of telemetry saturated data
and the image-center exclusion),
but constrained to be the same between HXD-PIN and HXD-GSO.
As the detector responses, we use 
\texttt{ae\_xi2\_20060213.rmf} and 
\texttt{ae\_xi2\_xisnom4\_20060615.arf} for XIS2,
\texttt{ae\_hxd\_pinxinom\_20060814.rsp} for PIN,
while \texttt{ae\_hxd\_gsoxinom\_20060321.rsp} 
and \texttt{ae\_hxd\_gsoxinom\_20070424.arf} for  GSO.
The last GSO file is the correction factor described in Paper~I,
introduced so as to make the Crab Nebula spectrum
described by a single PL model with  $\Gamma = 2.1$
\footnote{http://www.astro.isas.jaxa.jp/suzaku/analysis/hxd/gsoarf/}.
We discard the XIS data in the 1.7--1.9 keV range
to avoid complexity of the Si-K edge modeling,
and those in the $>8$ keV band 
which is possibly affected still by the photon pile-up effects.
The HXD-PIN data below 12 keV are also discarded in the fitting,
to avoid response uncertainties there \citep{KokubunHXD}.
We utilize the HXD-GSO data in the 70--400 keV energy band,
because of the poor signal significance (figure~\ref{fig:avspec})
and relatively large response uncertainties,
in $>400$ keV and $<70$ keV, respectively.
We assign a systematic error of 1\% to all the spectral bins,
in order to absorb uncertainties in the instrumental calibration,
including in particular those associated with the detector responses
and background estimation.
The effect of contamination on the optical blocking filters
of the XIS is modeled as additional absorption \citep{KoyamaXIS}.
We also trimmed the energy offset by $\sim$ 17 eV,
considering the current uncertainty 
with the XIS energy scale in the $2\times2$ mode.

\begin{figure}[bht]
\centerline{
\FigureFile(8cm,){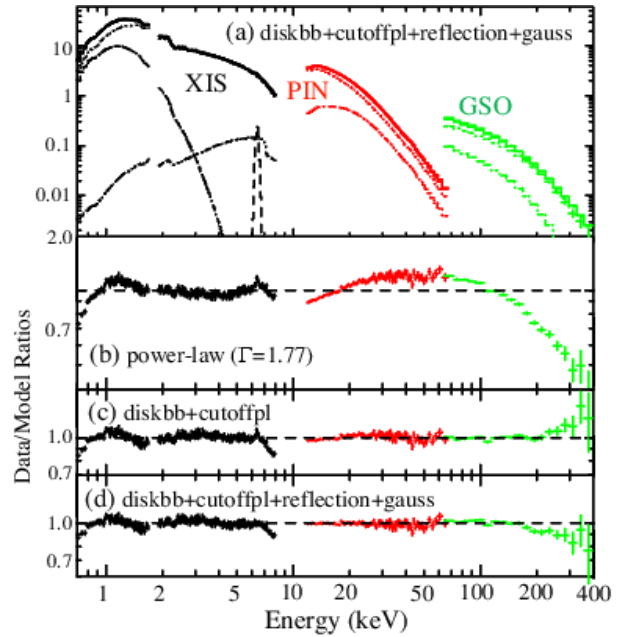}
}
\caption{Simultaneous fittings to the time-averaged XIS2,
HXD-PIN, and HXD-GSO spectra of Cyg X-1,
using various combinations of conventional models.
(a) Background-subtracted and response-inclusive spectra
(the same as in figure~\ref{fig:avspec}),
compared with the best-fit model consisting of 
a disk-blackbody, a cutoff power-law, 
a Gaussian, and a reflection component;
i.e., \texttt{wabs*(diskbb+cutoffpl+pexrav+gau)}.
Individual model components are separately drawn.
(b) Data-to-model ratios for an absorbed single-PL fit.
(c) The same as panel b, but a disk-blackbody component is added,
  and the power-law is replaced by a cutoff power-law model.
(d) The data-to-model ratios corresponding to the fit in panel a,
 namely, obtained by further adding to the model of panel c
 a Gaussian and a reflection component.
}
\label{fig:convfits}
\end{figure}

As the simplest attempt, 
we first fitted an absorbed single PL model
simultaneously to the three sets of spectra.
The fit indicated  $\Gamma \sim 1.77$ as the best estimate,
and an absorbing hydrogen column density of 
$N_{\rm H} = 3.5 \times 10^{21}$ cm$^{-2}$.
However, the fit was far from  acceptable,
with $\chi^2/\nu = 14.3$ for $\nu = 360$ degrees of freedom.
As seen in figure~\ref{fig:convfits}b,
the fit leaves several noticeable discrepancies,
including  a prominent data deficit above $\sim 100$ keV,
and a  data excess in 0.9--1.5 keV.
Obviously,
the former is due to the high-energy cutoff,
while the latter is due to the presence of  soft excess,
both of which have been established through previous observations.

To express  the high-energy turn over,
we next replaced the PL component with a cutoff power-law  model 
(\texttt{cuoffpl}; a power-law times an exponential factor).
To account for the soft excess,
we also added a disk-blackbody component
(\texttt{diskbb}: \cite{Mitsudadiskbb,Makishima1986})
to our fitting model.
The \texttt{diskbb} model is a simple non-relativisitic approximation
to the X-ray spectrum integrated over a standard accretion disk,
and is considered to provide an  approximately 
(within $\sim 10\%$; \cite{Watarai2000}) correct innermost disk radius  
when an appropriate correction factor is incorporated \citep{Aya1998}. 
Thus, the model is  \texttt{wabs*(diskbb+cutoffpl)} in the xspec  terminology;
here, we utilize version 11 of the xspec spectral fitting package
\footnote{ttp://www.heasarc.gsfc.gov/docs/xanadu/},
together with the photoelectric absorption model, \texttt{wabs}, by \citet{wabs}.
Then, the fit was improved drastically  to $\chi^2/\nu = 3.7$ for $\nu = 357$,
though not yet acceptable.
The exponential cutoff energy was obtained as $T_{\rm cut} \sim 170$ keV,
and the nominal PL photon index decreased to $\Gamma \sim 1.45$.
The innermost disk temperature of the \texttt{diskbb} component
was obtained as $T_{\rm in} \sim$ 0.5 keV,
and its innermost radius as  $r_{\rm in} \sim 13$ km 
at a distance of 2.5 kpc (section~\ref{sec:intro}),
without any correction for the color hardening factor,
or inner boundary condition,  or  inclination (see \cite{Makishima2000}).
We regard these \texttt{diskbb} parameters still unphysical.

In figure~\ref{fig:convfits}c, 
the data to model ratios from the \texttt{wabs*(diskbb+cutoffpl)} fit
show a slight broad hump over 20--40 keV
(though not very clear in this presentation),
indicative of reflection from a cool thick material.
In addition, there is clear evidence of Fe-K line at about 6.4 keV.
Accordingly, we added a reflection component (\texttt{pexrav}; \cite{MZ1995})
and a Gaussian to our model,
thus constructing a model \texttt{wabs*(diskbb+cutoffpl+pexrav+gau)},
and repeated the fitting.
In  using \texttt{pexrav},
the reflector was assumed to have an inclination $i=45^\circ$,
and its incident photons to have the same spectral shape
and normalization as those of the \texttt{cutoffpl} continuum
(\texttt{rel\_refl}$ < 0$ technically).
The Gaussian was allowed to have a free centroid  and a free width.
Then, the fit further improved to $\chi^2/\nu = 2.4$ for $\nu = 353$,
implying that the reflection is significantly present as reported by 
many previous observations.
The model gave the \texttt{cuoffpl} parameters as 
$\Gamma \sim 1.74$ and $T_{\rm cut} > 1000$ keV,
and the \texttt{diskbb} parameters as 
$T_{\rm in} \sim 0.3$ keV and $r_{\rm in} \sim 45$ km.
The reflector was found to have a moderate solid angle as 
$\Omega/2\pi \sim 0.5$,
and the Gaussian centroid and its width were obtained 
as $6.43$ keV and $\sigma < 0.2$ keV respectively.
In figure~\ref{fig:convfits}a, we show this fit 
where contributions of 
the individual model components are separately drawn.
The data-to-model ratios are given in figure~\ref{fig:convfits}d.
Thus, the reflection signal is inferred to carry
$\sim$ 30\% of the continuum at 30 keV.

In this way,
the empirical model fittings have roughly quantified 
the time-averaged Suzaku spectra of Cyg X-1,
and extracted their basic features.
The fit implies a 0.7--300 keV flux (after absorption by $N_{\rm H}$) 
of $5.4 \times 10^{-8}$ erg cm$^{-2}$ s$^{-1}$.
Assuming an isotropic emission,
and employing  the distance of 2.5 kpc (section~\ref{sec:intro}),
the absorption-removed  0.7--300  keV luminosity 
becomes  $4.6 \times 10^{37}$ erg s$^{-1}$,
which is typical of this object.
Although the fit still remains unacceptable,
and hence errors associated with the model parameters cannot be evaluated,
further  improving  the  model would be of little meaning,
because \texttt{cutoffpl}  is a mere empirical approximation
to the spectra expected from unsaturated Comptonization processes.
Instead, we  proceed to more physical modelings,
using theoretical Comptonization codes.

\subsection{Fitting with Comptonization models
\label{subsec:comppsfits}}

Among various Comptonization codes currently available,
here we employ that of Poutanen \& Svensson (1996),
called \texttt{compPS} in xspec,
because of the following three advantages:
it correctly takes into account the relativistic effects
in the Compton scattering kinematics;
it allows us to use a \texttt{diskbb} model as the seed photon source;
and it has a built-in function to allow a part of the initial 
Compton-produced photons to get into a cold matter and reflected.

We thus fitted the XIS2, HXD-PIN, and HXD-GSO spectra 
simultaneously by a \texttt{wabs*(diskbb+compPS)} model,
in the same manner as before.
At this stage,
the reflection option in \texttt{compPS} was suppressed,
and the Gaussian component was not included.
Here and hereafter, we  assume 
that the Compton cloud has a spherical geometry (parameter=4)
with an electron temperature $T_{\rm e}$ and an optical depth $\tau$,
and that the seed photons to the  \texttt{compPS} component are 
supplied by a cool standard disk having the same $T_{\rm in}$ 
(but a separate normalization) as the \texttt{diskbb} component.
The implied picture is  that some fraction of photons 
from a standard disk is fed to the Comptonizing corona,
while the rest is directly visible as the  spectral soft excess.
This picture is supported by the presence of a tight correlation
between the continuum slope and the reflection strength 
in BHBs \citep{ZLS1999,Gilfanov1999,Revnivtsev2001} and in AGNs,
which indicates that the optically-thick disk, producing the reflection,
is also causally related to the Comptonized emission.

\begin{figure}[bth]
\centerline{
\FigureFile(8cm,){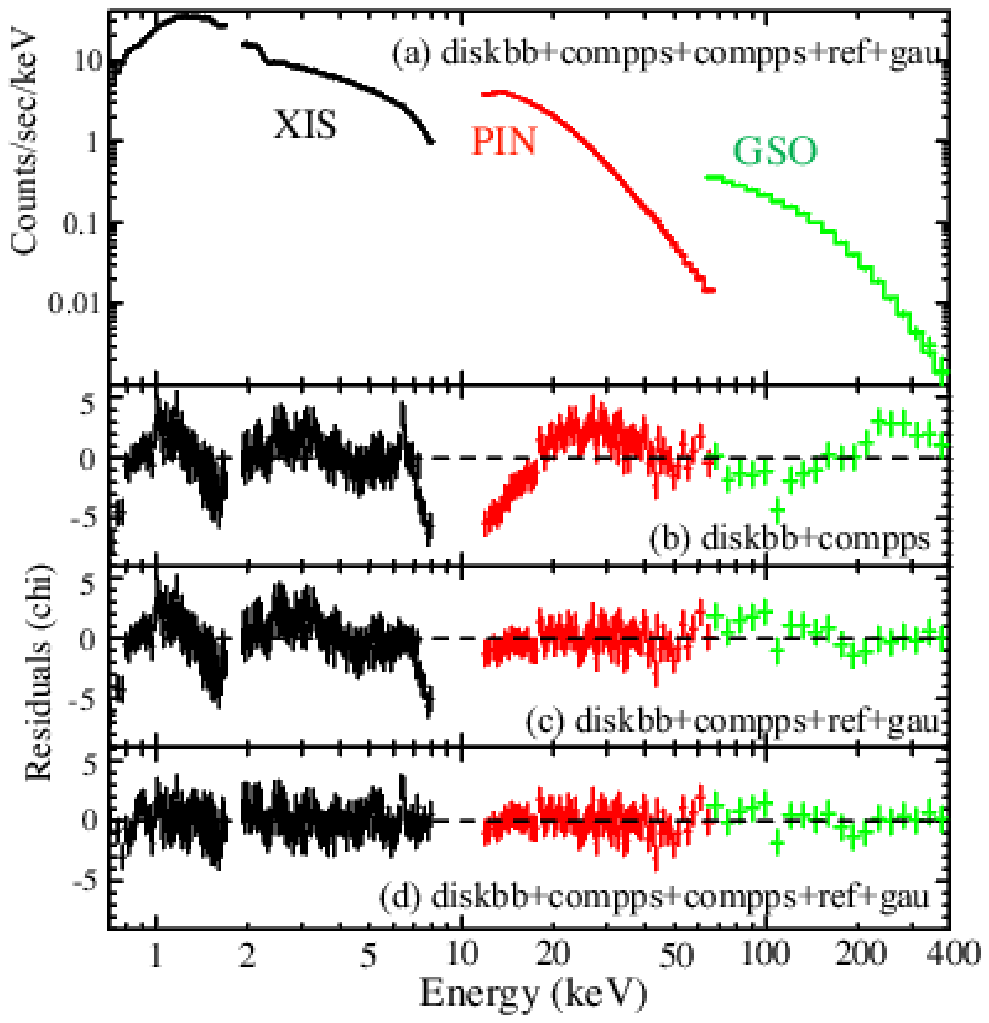}
}
\caption{Simultaneous fittings to the time-averaged XIS2,
HXD-PN, and HXD-GSO spectra of Cyg X-1,
employing thermal Comptonization models.
The derived best-fit parameters are given in table~\ref{tbl:parameters}.
(a) The same spectra as in figure~\ref{fig:convfits}a,
compared with the best-fit \texttt{wabs*(diskbb+compPS+compPS+gau)} model.
The two \texttt{compPS} components are constrained
to have the same $T_{\rm e}$, 
and a common set of reflection parameters.
(b) Fit residuals from a model incorporating a single \texttt{compPS} continuum
without reflection, together with a \texttt{diskbb}.
(c) The same as panel b, but reflection is incorporated,
and a Gaussian for the Fe-K line is added.
(d) The residuals correspond to the fit shown in panel a.
Compared to panel c, the second \texttt{compPS} component is  added.
}
\label{fig:comppsfits}
\end{figure}

As shown in figure~\ref{fig:comppsfits}b, 
this model gave a poor fit with $\chi^2/\nu = 3.9$ for $\nu = 357$.
Since figure~\ref{fig:comppsfits}b 
reveals  the reflection hump at $\sim 30$ keV,
we activated the reflection option in \texttt{compPS}.
This has nearly the same effect as the \texttt{pexrav} model,
but with two additional free parameters;
 the ionization parameter $\xi$ in units of erg cm s$^{-1}$,
and the innermost reflector radius $R_{\rm in}^{\rm ref}$.
The latter controls relativistic blurring of spectral features,
and is expressed in units of the gravitational radius $R_{\rm g} = GM_{\rm BH}/c^2$,
with $M_{\rm BH}$ the BH mass and $G$ the gravitational constant.
The outer radius was  fixed at $10^{6} R_{\rm g}$ 
and the radial emissivity profile was assumed to
depend on the radius $r$ as $r^{-3}$ \citep{Fabian89}.
The  inclination angle, abundance, and temperature of the reflector
were also fixed at $45^{\circ}$, 1 solar, and $10^6$ K, respectively.
We retained the Gaussian component,
because the \texttt{compPS} reflection does not include any emission line.
Then, as shown in figure~\ref{fig:comppsfits}c,
the fit was much improved ($\chi^2/\nu = 2.6$ with $\nu = 351$);
the reflection hump is significantly present with $\Omega/2\pi \sim 0.2$ and $\xi \sim 0$,
although  $R_{\rm in}^{\rm ref}$ was not constrained in this and all subsequent fits.
The fit goodness is comparable to that obtained 
with the \texttt{wabs*(diskbb+cutoffpl+pexrav+gau)} model (figure~\ref{fig:convfits}d).
While the conventional model failed to constrain  $T_{\rm cut}$,
the present model gave an electron temperature of $T_{\rm e} = 115$ keV.

In figure~\ref{fig:comppsfits}c,
the fit  still leaves a prominent data deficit in the highest XIS range.
In addition, the XIS vs. HXD normalization ratio was obtained at 0.108,
which  exceeds  the nominal value of $\sim$ 0.092 (subsection~\ref{subsec:xisreduction});
this is too large,
even considering uncertainties in the correction factors 
involved in the XIS and HXD data processing  (section~\ref{sec:obs}).
We hence suspect
that the actual source continuum is rising from 
$\sim 10$ keV toward lower energies, 
with a steeper slope than indicated by the model,
so that the fitting algorithm tried to absorb this mismatch
by artificially raising the  model prediction to the XIS data.
In other words, 
the broad-band continuum has a more concave shape 
over 2--20 keV  (see figure~\ref{fig:convfits}b),
than would be explained by the \texttt{compPS} component with reflection.
This effect was previously noticed by \citet{DiSalvo2001} as a soft excess 
that cannot be accounted for by the cool disk emission.
The same effect was noticed in the Suzaku 
data of GRO~J1655$-$40 (Paper~I),
and led \citet{Takahashi1655}
to incorporate another \texttt{compPS} component.
Such a ``double-Compton'' modeling was previously applied successfully 
to the LHS of Cyg X-1 in several works;
by \citet{G1997} and  \citet{Ibragimov2005} using  simultaneous Ginga,
RXTE, and CGOR/OSSE data,
and by  \citet{Frontera2001} to  the BeppoSAX  data.

We hence tried a model incorporating two Comptonized continua,
which are constrained to have the same seed-photon temperature
(which in turn is the same as that of \texttt{diskbb}),
the same hot-electron temperature,
and the same reflection parameters,
but are allowed to differ in  their normalization and optical depth.
The model is hence \texttt{wabs*(diskbb+compPS+compPS+gau)}.
As shown in figure~\ref{fig:comppsfits}d,
the fit has been improved drastically
to $\chi^2/\nu = 1.13$ ($\nu = 349$); 
the XIS, PIN, and GSO spectra are reproduced
with $\chi^2/\nu=$ 1.3, 0.9, and 0.9, respectively.
The model has  removed the data deficit in the highest XIS range,
and reduced a residual structure near 2 keV
where the two \texttt{compPS} components now cross over.
Furthermore, the XIS vs. HXD normalization has became 0.088,
which is close to the value of 0.092
calibrated using SS Cyg (subsection~\ref{subsec:xisreduction}).
We are therefore confident that the Suzaku data  
require two (or possibly more) Compton components.
Although the fit is not yet formally acceptable, 
we regard it as satisfactory,
because the data vs. model discrepancy,
typically within 4\% over the entire 0.7--400 keV range,
is comparable to those obtained 
when we fit the Suzaku spectra of bright objects, 
such as the Crab spectra,  with simple models.

\begin{figure}[bht]
\centerline{
\FigureFile(7.5cm,){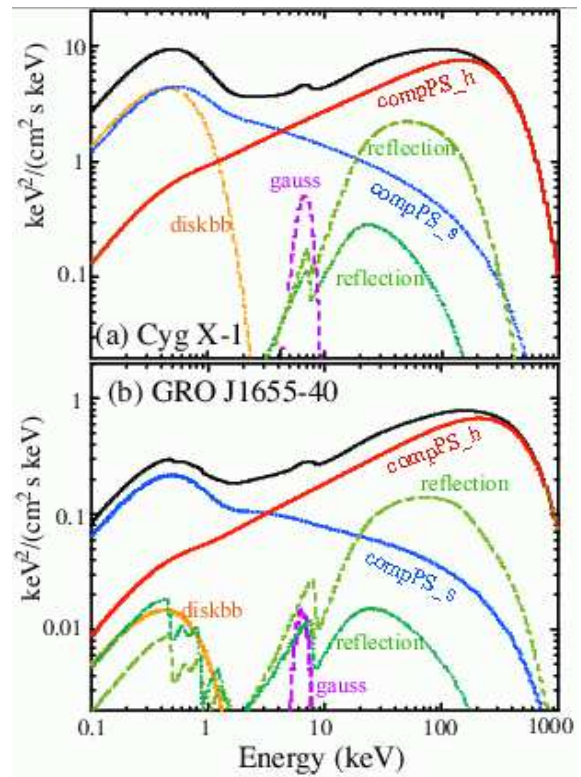}
}
\caption{(a) The inferred best-fit $\nu F \nu$ spectrum of Cyg X-1,
shown with the absorption removed.
It corresponds to panels a and e of figure~\ref{fig:comppsfits},
with the model parameters detailed  in the  column labeled ``Average''
of table~\ref{tbl:parameters}.
Red, blue, orange,  green,  and purple specify
\texttt{compPSh}, \texttt{compPSs}, \texttt{diskbb},  reflection components,
and a Gaussian for the Fe-K line, respectively.
(b) The same as panel a, 
but for GRO~J1655$-$40 observed with Suzaku on 2005 September 22 and 23.
It is identical to figure 7  of Paper~I,
except for the removal of the absorption.}
\label{fig:bestfitmodels}
\end{figure}

Our final model obtained in this way is shown 
in figure~\ref{fig:comppsfits}a in the convolved form,
and in figure~\ref{fig:bestfitmodels}a in the incident $\nu F \nu$ form.
Its model parameters are given in table~\ref{tbl:parameters}.
Hereafter, we call the \texttt{compPS} components
with the larger and smaller $y$-parameters
\texttt{compPSh} and \texttt{compPSs}, respectively,
with the suffix ``h'' standing for ``hard''  and ``s'' for ``soft''.
In figure~\ref{fig:bestfitmodels}, the \texttt{compPSs} component
includes not only the scattered photons [$\propto 1-\exp(-\tau)$],
but also those seed photons which traversed the
Compton cloud without scattered  [$\propto \exp(-\tau)$].

\subsection{Examination of the Comptonization model results
\label{subsec:comppsfits_examination}}
Although the analysis conducted in the preceding subsection 
has provided a plausible representation of the time-averaged spectra,
different spectral models could often degenerate.
Therefore, let us briefly examine our final spectral model 
for its implications and physical consistency,
leaving detailed discussion  to section~\ref{sec:discussion}.

The obtained absorption,
$(6.6^{+0.8}_{-0.3}) \times 10^{21}$ cm$^{-2}$,
is  close to  the value of $(5.3 \pm 0.2) \times 10^{21}$ cm$^{-2}$
derived by \citet{Dotani1997} with ASCA.
The \texttt{compPSh} and \texttt{compPSs} components, 
crossing at $\sim 4$ keV, 
are characterized by photon indices of $\sim 1.6$ and $\sim 2.4$, respectively,
with the former consistent with the approximate X-ray  slope of Cyg X-1
measured in energies of a few to a few tens keV.
The underlying parameters,
$T_{\rm e} \sim 100$ keV, 
$\tau$ of order unity, and $\Omega/2\pi \sim 0.4$,
are consistent with a large number of past measurements 
(e.g., \cite{G1997, DiSalvo2001,Frontera2001,ZG2004,Ibragimov2005}).

The \texttt{diskbb} component has a raw innermost radius 
as $r_{\rm in} = 180^{+150}_{-50}$ km,
which yields the best-estimate actual inner radius of 
$R_{\rm in} =250^{+210}_{-60}$ km;
this is obtained by  applying the nominal correction factors in \citet{Makishima2000} 
(including a color hardening factor of 1.7; \cite{ShimuraTakahara}) to $r_{\rm in}$,
and  correcting for the inclination assuming  $i=45^\circ$, as
\begin{equation}
R_{\rm in} = 1.18 r_{\rm in} /\sqrt{\cos i}~.
\label{eq:rin2Rin}
\end{equation}
This $R_{\rm in}$ is  larger than  the value of $R_{\rm in} \sim 100$ km
measured in the high/soft state  with ASCA 
\citep{Dotani1997,Makishima2000},
suggesting a disk truncation at a larger radius
than in the high/soft state (see also subsection~\ref{subsec:diskpara}).
The innermost disk temperature, 0.19 keV,
is a factor 2.3 lower than that measured in the high/soft state \citep{Dotani1997},
consistent with a reduced accretion rate
and possibly a larger innermost disk radius.
These consistencies support (though not necessarily prove) our basic assumption
that the seed photons for Comptonization are provided by the optically-thick disk,
as proposed previously \citep{ZLS1999,Gilfanov1999,Revnivtsev2001}.
Although the bolometric luminosity of the directly-visible 
disk emission is $19^{+24}_{-6}$\% of the overall 0.7--300 keV luminosity,
the energy generation in the cool disk is approximately twice higher,
when the seed photon luminosity is added.
Then, the energy generation in the the cool disk
becomes comparable to  that in the hot corona.

While the \texttt{compPSh} component carries the dominant hard continuum,
\texttt{compPSs}  successfully explains  the continuum  steepening 
in energies below a few keV,
or equivalently, the soft excess that is too hot to be explained
by the cool disk emission \citep{DiSalvo2001}.
These two Compton continua are inferred
to have optical depths which differ by a factor of $3.9 \pm 0.5$.
Their normalizations,
originally expressed as the  disk area $A$ supplying the seed photons,
have  been converted in table~\ref{tbl:parameters}
to its innermost radius $R_{\rm in}^{\rm seed}$ as
\begin{equation}
     R_{\rm in}^{\rm seed} = 1.18 \sqrt{A} ~.
\label{eq:seeddisk}
\end{equation}
Though similar to $R_{\rm in}$ of equation~(\ref{eq:rin2Rin}), 
$R_{\rm in}^{\rm seed}$  is {\em not} corrected for the inclination,
since  we presume the Comptonized emission to be approximately isotropic.
As seen in table~\ref{tbl:parameters},
the value of  $R_{\rm in}^{\rm seed}$ associated with \texttt{compPSs}
is comparable to  $R_{\rm in}$,
while that of \texttt{compPSh} is considerably smaller.
We hence obtain an implication 
that $\sim 7$ times more seed photons are fed to \texttt{compPSs}
than to \texttt{compPSh}:
a further discussion continues in subsection~\ref{subsec:diskpara}.

 Although the final model is generally successful,
the obtained XIS vs. HXD normalization,
0.088,  is  $\sim 5\%$ too small.
When  this ratio is fixed to the calibration value of 0.092,
the PIN data exhibit some positive and negative residuals 
at $\sim 15$ keV and  $\sim 40$ keV, respectively.
These can be eliminated by allowing  \texttt{compPSs} and \texttt{compPSh} 
to have a larger ($\sim 1.0$) and a slightly smaller ($\sim 0.3$) 
value of $\Omega/2\pi$, respectively,
with the raw chi-square reducing by 13 (for $\nu=349$).
This is because the reflection associated with \texttt{compPSs}
appears at lower energies than those with \texttt{compPSh}
(figure~\ref{fig:bestfitmodels}a).
There is thus some indication
that the softer Compton signals are more strongly reflected
by cold materials than the harder one.

When the two \texttt{compPS} components 
are allowed to take different values of $T_{\rm e}$,
tolerances of the model parameters  increase,
but the fit goodness remains the same.
This is because the high-energy roll over of \texttt{compPSs} is masked by \texttt{compPSh}.
Therefore, we do not adopt this modeling,
although the data do not necessarily exclude the presence of multiple electron temperatures.


\section{Analysis of Spectral Changes Associated with the Random Variation}
\label{sec:ana_varspec}

\subsection{Methods}
\label{subsec:varspec_method}
Now that the time-averaged spectra have  been 
quantified in section~\ref{sec:ana_avspec},
we proceed to the latter half of our data analysis,
namely attempts to understand  how the spectrum
changes as the source varies  randomly;
a similar attempt was carried out previously by \citet{TID2001}.
For this purpose,
it is essential to sort the broad-band spectra,
particularly those of the HXD, into two subsets,
according to whether the instantaneous source count rate 
is higher or lower than a certain mean value.
However, the HXD has a relatively small effective area
($\sim 230$ cm$^2$ at 100 keV; \cite{TakahashiHXD}),
so the average count rates of Cyg X-1
are  $\sim 47 $ c s$^{-1}$ and $\sim 20 $ c s$^{-1}$
in the 10--70 keV (with PIN) and 70--400 keV (with GSO) bands,
respectively (subsection~\ref{subsec:obs}; figure~\ref{fig:ltcurves}c,d).
Then, even in the PIN range where the NXB is relatively minor,
Poisson fluctuations of signal counts
with 1 s integration would  become comparable 
to the intrinsic variations of  Cyg X-1
($\sim \pm 20$\% in root-mean-square amplitude).
Making the integration time longer would not help,
because  Cyg X-1 varies with an approximately 
``white'' power spectrum on time scales longer than a few seconds
(e.g., \cite{BH1990}),
and hence amplitudes of the intrinsic and Poisson variations 
will scale in a similar manner 
when the integration time is increased. 

Fortunately, we can utilize the XIS for our purpose,
because one XIS sensor received $\sim 60 $ c s$^{-1}$ 
even after the image center has been removed to avoid 
photon pile up (subsection \ref{subsec:xisreduction}; figure~\ref{fig:ltcurves}a,b),
and the variation of Cyg X-1 is known to be crudely synchronized
across the soft X-ray to hard X-ray energies \citep{NKM1994}.
Since the present XIS data have a 1 s time resolution (subsection \ref{subsec:obs}),
we can  judge every second, referring to the XIS data, 
whether the source is brighter or fainter than a mean,
and sort the XIS and HXD data into the required two subsets.
In practice, we have constructed the following procedure to do this.

\begin{enumerate}
\item 
In the same way as in section~\ref{subsec:obs_ltcurve},
a 0.3--10 keV  XIS0+XIS3 light curve $\{C_i\}$ 
(with $i$ the bin number) is produced throughout the observation,
with a bin width of  $t_{\rm b} \geq 1$ s.

\item
The calculation of  $\{C_i\}$ excludes those 1-s time bins 
in which either XIS0 or XIS3 (or both) suffered  telemetry saturation.

\item
Every $T$ s ($T \gg t_{\rm b}$), 
a  mean XIS0+XIS3 count rate $\bar{C}$
is calculated from $\{C_i\}$.

\item
Two sets of good time intervals (GTIs),
called ``high-phase'' (HP)  and ``low-phase'' (LP),  
are defined, by judging every $t_{\rm b}$ s  
whether  $C_i$ is higher than  $\bar{C}  + n\sigma$,
or lower than $\bar{C}- n\sigma$, respectively.
Here, $\sigma$ is 1-sigma Poisson error
and $n \geq 0$ is a discriminator level.

\item
Using the same two GTIs,  the XIS2 data are accumulated into two spectra, 
called HP (high phase) and LP (low phase) spectra.
The data are discarded  if $\bar{C}-n\sigma<  C_i < \bar{C}+n\sigma$.

\item
Employing the same two GTIs,
the  HXD-PIN data are sorted into HP and LP spectra;
so are the HXD-GSO data.
 
\item
Using the two sets of GTIs,
two  background data sets are accumulated for  HP and LP,
and are subtracted from the corresponding on-source HXD spectra.

\end{enumerate}

Thus, the procedure involves three adjustable parameters,
$t_{\rm b}$, $T$, and $n$.
Among them, $T$ should be long enough to suppress Poisson errors,
but short enough to compensate for background changes.
The threshold  $n$ should also be optimized,
because a larger $n$ leads to a decrease of live time,
while a smaller $n$ means that the data sorting is
diluted by the Poisson fluctuation in $\{C_i\}$
which is not related to the intrinsic source variation.
Here, for simplicity, we have chosen  
$t_{\rm b}=1$ s (the highest time resolution of the XIS),
$T=200$ s, and $n=0$ as our baseline conditions.
This is because a larger value of $n$, such as $n=1$,
was found to enhances the HP vs. LP ratio,
but also to degrade the data statistics,
thus hampering us to more accurately
separate the HP and LP model parameters.
Since we have chosen to use the  XIS0+XIS3 data  for the intensity judgement,
we derive the intensity-sorted XIS spectra 
only from the other FI sensor, namely XIS2.
This division of task  is needed
because the intensity-sorted spectra
from XIS0 and XIS3 would include  those time bins
when their summed counts are higher (or lower) 
due simply to statistical fluctuations,
which amount to $\sim 8\%$ (1 sigma) at the 1-s integration.
The intensity-sorted spectra from XIS0 and XIS3,
if to be utilized in the spectral fitting,
would need to be corrected for this effect.
The XIS2 data, in contrast, are free from this problem,
because Poisson errors from different detectors are independent.

\subsection{Auto-correlation and cross-correlation functions}
\label{subsec:acf_ccf}

Before actually conducting the analysis described 
in section~\ref{subsec:varspec_method},
we briefly  perform auto-correlation function (ACF)
and cross-correlation function (CCF) analyses,
in order to examine
whether the GTIs sorted in reference to the energy band below 10 keV
is really applicable to the extraction of 
varying spectral components in harder energies.
We utilize raw light curves with neither background subtraction
nor dead-time correction.

\begin{figure}[hbt]
\begin{center}
\FigureFile(16cm,){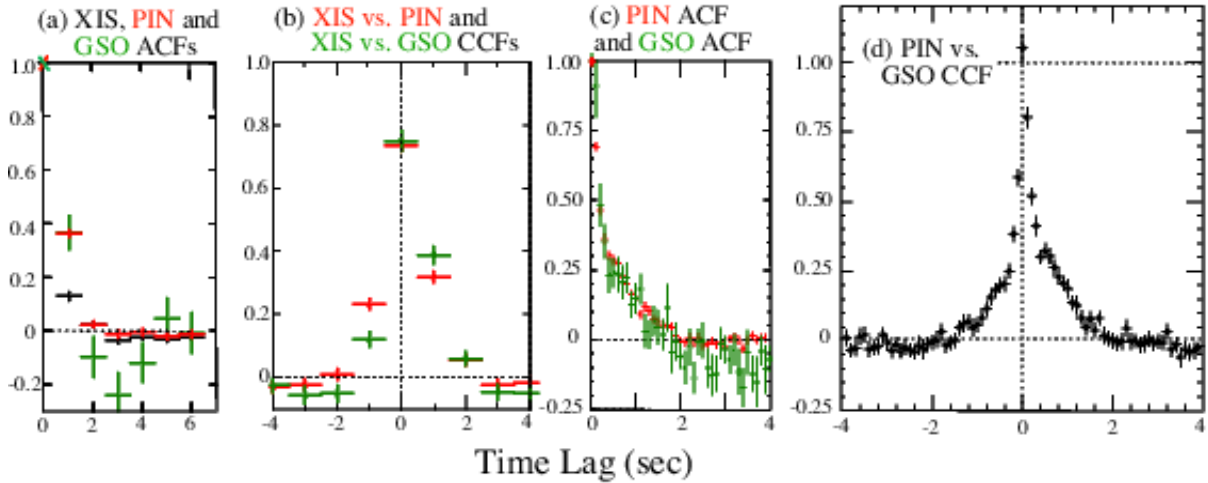}
 \end{center}
\caption{Auto-correlation functions (ACFs) 
and cross-correlation functions (CCFs)
of the present Suzaku data.
The effects of Poisson errors were removed,
and the CCFs were normalized so that the value at zero lag
directly represent the correlation coefficient.
Positive time differences in CCFs mean 
that softer signals lead the harder ones.
(a) ACFs from the XIS2 (0.7--10 keV; black),
HXD-PIN (10--60 keV; red), and HXD-GSO (60--200 keV; green) data,
calculated using  1-s binned light curves.
(b) CCFs of the PIN (red) and GSO (green) data,
calculated against the 0.7--10 keV
XIS2 light curve in the same way as panel a.
(c) ACFs calculated using 0.1-s binned 
PIN (10--60 keV;  red) and GSO (60--200 keV; green) light curves.
(d) CCFs of the GSO
light curve against that of  PIN,
calculated in the same way as panel c.
}
\label{fig:acf_ccf}
\end{figure}

Panels (a) and (b) of figure~\ref{fig:acf_ccf}  show 
ACFs and CCFs from the present Suzaku data, respectively,  
calculated employing a  bin width of  1 s.
They were calculated
using data intervals each 256 s long,
and then averaged over ensemble.
The XIS telemetry saturation was treated in the same way as
item  2 of the previous subsection (using $t_{\rm b}=1$ s).
Thus, the XIS vs. PIN and XIS vs. GSO CCFs 
both exhibit clear peaks at zero time lag,
with correlation coefficients of $\sim 0.75$.
This ensures that the intrinsic variation on a time scale of 1 s
is well correlated over the entire energy range 
in which we are conducting our analysis,
and validate our method described in the previous subsection.

Although the XIS data do not have time resolution higher than 1 s,
the HXD data afford it down to $61~\mu$s 
\citep{TakahashiHXD, TeradaHXDtiming}.
We calculated  ACFs and CCFs  from the 
PIN (10--60 keV) and GSO (60--200 keV) light curves accordingly,
with 0.1 s resolution and 409.6 s  data intervals.
The results,  shown in panels (c) and (d) of figure~\ref{fig:acf_ccf}, 
reveal that the variations are detected on shorter time scales,
where the PIN vs GSO correlation is still peaked at zero-lag.
The variations in the two energy bands are again well synchronized, 
with the correlation coefficient of $\sim 1$.

In addition to the basic properties described above,
these ACFs and CCFs reveal one interesting property:
namely, the left-to-right asymmetry in the CCFs,
seen on both time scales,
in such a way that the correlation is stronger toward hard-lag sense.
This reconfirms the previous report of the same phenomenon
in early days by  \citet{Noloan1981},
then  using Ginga by \citet{Miyamoto1991},
and further with RXTE (e.g., \cite{Nowak1999a,MCP2000}).
On the other hand,
the present results do not clearly reconform another property,
reported  previously (e.g., \cite{NKM1994,FLC1999,MCP2000}),
that  ACFs in softer energies have wider correlation peaks,
and hence longer correlation lengths.
 We leave this issue to future Suzaku observations using higher XIS time resolution
(employing so-called PSUM mode);
then, we will also be able to study how the spectrum differs
between the rising and falling phases of ``shots",
as suggested previously \citep{NKM1994}.

\onecolumn

\begin{figure}[hbt]
\begin{center}
\FigureFile(14cm,){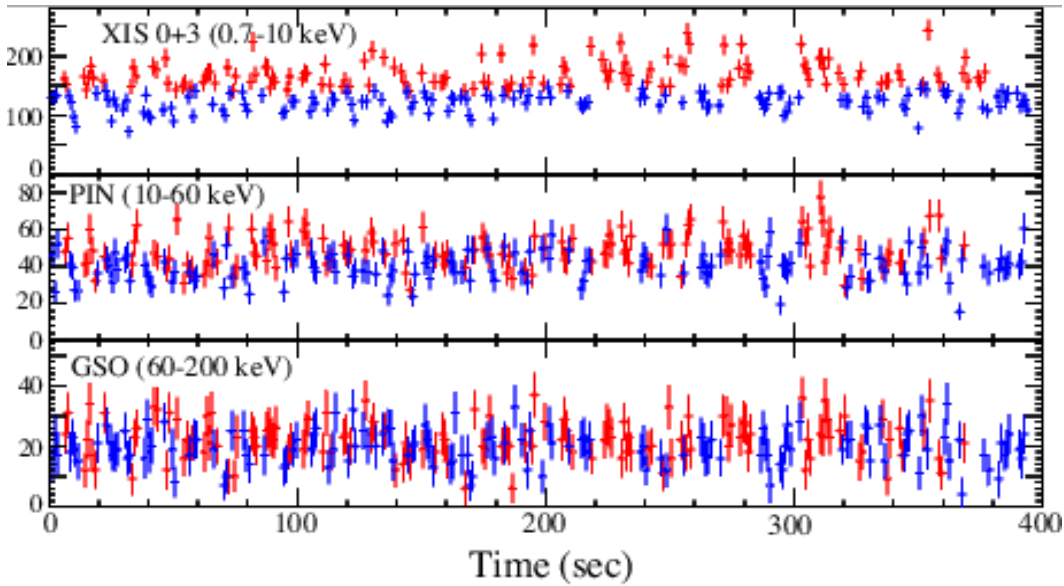}
 \end{center}
\caption{Background-inclusive  XIS (top: 0.7--10 keV),
 HXD-PIN (middle: 10--60 keV), and HXD-GSO (bottm: 60--200 keV)
 light curves of Cyg X-1 with 1-s binning,
 sorted according to the instantaneous XIS0+XIS3 counts.
 Red and blue data bins represent those when the XIS0+XIS3 counts, $C_i$,
 are higher and lower than the local 200 s average, $\bar{C}$, respectively.
 No correction is made for the instrumental dead times.}
\label{fig:ltcurve_1sec}
\end{figure}

\twocolumn
\subsection{Intensity-sorted spectra}
\label{subsec:sortedspec}

Now that the examination of CCFs have validated 
the method presented in section~\ref{subsec:varspec_method},
we have applied it to the present Suzaku data.
Figure~\ref{fig:ltcurve_1sec} gives an example of 
1-s bin light curves of Cyg X-1,
sorted according to the above method using the baseline condition;
$t_{\rm b}=1$ s, $T=200$ s, and $n=0$
(section~\ref{subsec:varspec_method}).
In this particular example,
the average  XIS count rate used as the HP vs. LP  threshold
is 140 and 148 ct s$^{-1}$ in the former and latter 200 s, respectively.
These are $\sim 15\%$  higher than the XIS2 counting rate 
indicated by figure~\ref{fig:ltcurves} (panels a, b),
because the XIS2 optical axis is more offset from the 1/8-window centroid
than those of XIS0 and XIS3, causing a larger photon loss.
The XIS light curve reveals the well known flickering behavior of Cyg X-1,
while those of the HXD suffer  larger Poisson errors
so that the intensity sorting on a time scale of 1 s is impossible 
without the XIS data.

\begin{figure}[thb]
\begin{center}
\FigureFile(8cm,){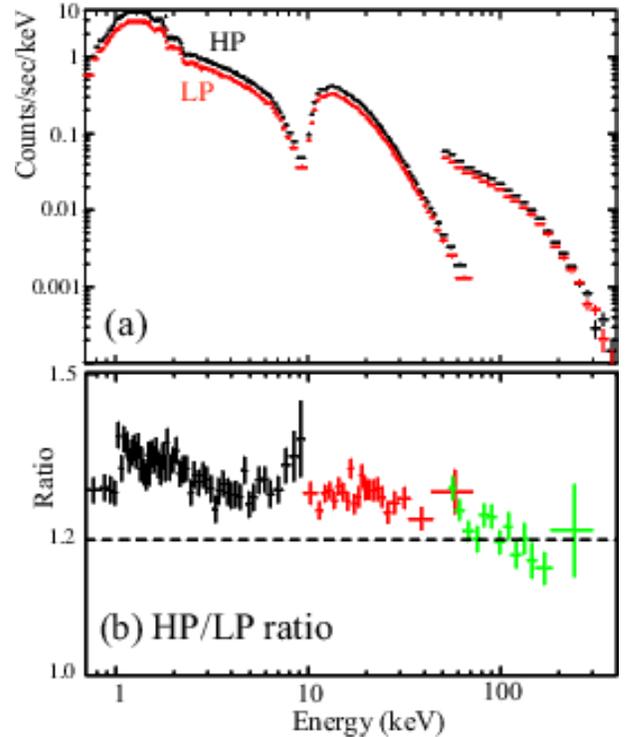}
 \end{center}
\caption{(a) The HP (high phase; black) and LP (low phase; red) 
spectra of Cyg X-1 recorded by XIS2, HXD-PIN, and HXD-GSO,
while XIS0+XIS3 data are used in the intensity judgement.
The corresponding HXD backgrounds have been subtracted.
(b) The ratios between the HP and LP spectra.
}
\label{fig:HiLo_spec}
\end{figure}

Figure~\ref{fig:HiLo_spec} shows the HP and LP spectra derived in this way.
There, we subtracted the  corresponding  backgrounds 
accumulated using the respective GTIs
(item 7 in section~\ref{subsec:varspec_method}):
the utilized two background sets agree  within their statistical errors.
While the spectra are presented without dead time correction,
we correct the model nomalizations (in table~\ref{tbl:parameters} and elsewhere in text)
for the time-averaged dead time of 7\%.
Although the HP and LP spectra should have different dead times,
the difference is calculated to be within $\sim 2\%$,
because the dead time is mainly caused by background counts.
This difference is expected to reduce 
the HP spectral normalization by at most  $2\%$,
keeping the spectral shape intact.

Figure~\ref{fig:HiLo_spec}b  gives the ratios between the HP and LP spectra,
with several important implications.
First, the average ratio becomes  $\sim 1.3$,
in agreement with the typical variation amplitude of Cyg X-1.
Second, the ratio increases at the highest XIS  energies band,
but this is due to still remaining photon pile-up effects in the XIS
when the source gets brighter, and should be ignored.
Third, the spectrum gets slightly softer when the source brightens up,
as evidenced by a gradual decrease of the HP/LP ratio toward higher energies;
this qualitatively agrees with previous results from, e.g.,  Ginga (\cite{NKM1994}).
Finally, the HP/LP ratio exhibits a rather abrupt decrease 
at $\sim 1$ keV toward lower energies.

In order to see the effects of changing the selection condition,
we repeated the intensity sorting using $t_{\rm b}=3$  s and 8 s.
The results obtained with $t_{\rm b}=3$  s were essentially the same 
as those with the baseline case ($t_{\rm b}=$ 1 s).
When  employing $t_{\rm b}=8$ s,
the  HP/LP ratio was reduced to $\sim 1.2$ on average,
although its energy dependence was very similar to the baseline case.
From these, we retain $t_{\rm b}=$ 1 s.
We also examined different Poisson thresholds, $n$.
When $n=0$ (the baselline case), 1, and 2 are used,
the HP GTIs  yielded an overall exposure of 7.3, 5.2, and 3.3 ks, respectively
(with LP GTI being comparable),
while the energy-averaged HP/LP ratio turned out to be
$\sim 1.3$, $\sim 1.4$, and $\sim 1.5$, respectively.
We  hence retain $n=0$, 
since it maximizes the signal-to-noise ratio 
of the HP minus LP difference spectra.

Returning to the HP and LP spectra obtained with the baseline condition,
we tried grossly quantifying them using the conventional model
employed in section~\ref{subsec:convfits}.
To avoid the complexity in softer energies 
which arisses from the iron line, the soft excess, and the softer Compnotization continuum,
we fitted only the HXD (PIN and GSO) spectra,
using the \texttt{cutoffpl+pexrav} model.
The model reproduced  the HP and LP spectra successfully both with $\chi/\nu \sim 1$,
and yielded the consistent reflection strength;
$\Omega/2\pi = 0.16^{+0.05}_{-0.06}$ in HP and $0.16 \pm 0.06$ in LP.
(These are smaller than the value of $\sim $0.4 obtained 
with the time-averaged spectra using the Compnonization models,
because a cutoff power-law model is more convex than an unsaturated Compton continuum.)

To highlight the difference between the two spectra,
we next  fitted them jointly,
with $\Omega/2\pi$ fixed at 0.16.
As show in table~\ref{tbl:HiLo_convfits} (column under an entry ``Free''),
we were unable to find statistically significant differences
either in $\Gamma$ or $E_{\rm cut}$.
Therefore, we further constrained the HP and LP spectra  
to have either the same $\Gamma$ or the same $E_{\rm cut}$.
Then, as shown in the last two columns of  table~\ref{tbl:HiLo_convfits},
the join fit remained acceptable
without preference between the two cases,
and the unconstrained parameter showed a statistically 
significant difference between the two spectra.
We therefore conclude that the HP and LP spectral have different shapes
as suggested by their ratio (figure~\ref{fig:HiLo_spec}b),
and that either $E_{\rm cut}$ or $\Gamma$ (or both) is changing,
although we cannot determine which is more variable.

\begin{figure}[bht]
\begin{center}
\FigureFile(7.5cm,){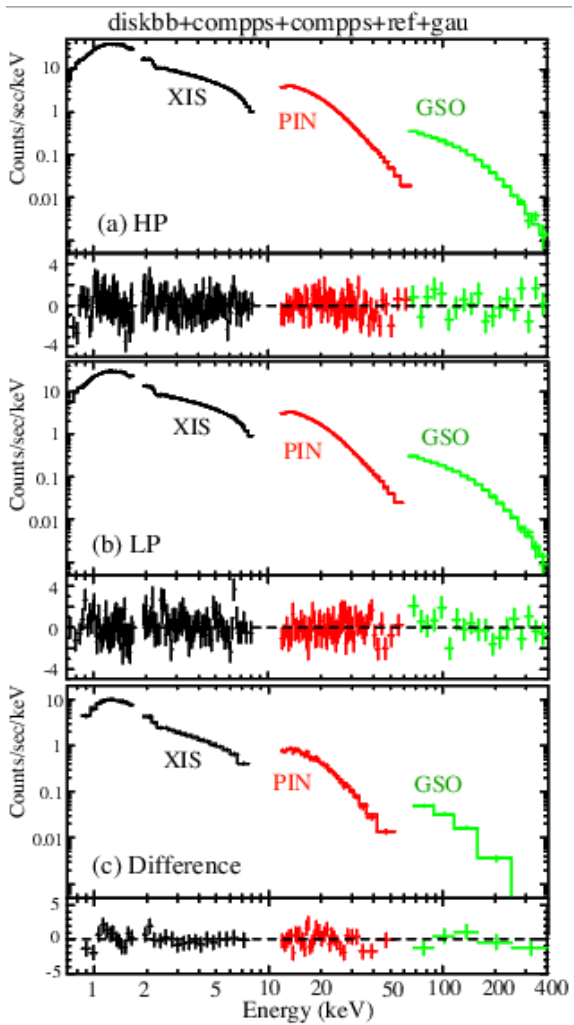}
 \end{center}
\caption{Background-subtracted and dead-time uncorrected
 HP (panel a) and LP (panel b) spectra of Cyg X-1,
 fitted with the \texttt{wabs*(diskbb+compPS+compPS+gau)} model
 in the same way as figure~\ref{fig:comppsfits}.
 Panel c is for the difference spectra between HP and LP.}
\label{fig:HiLo_compPSfits}
\end{figure}

\subsection{Changes of the Compton parameters}
\label{subsec:HiLo_compPSfits}
\begin{figure}[hbt]
\begin{center}
\FigureFile(7.5cm,){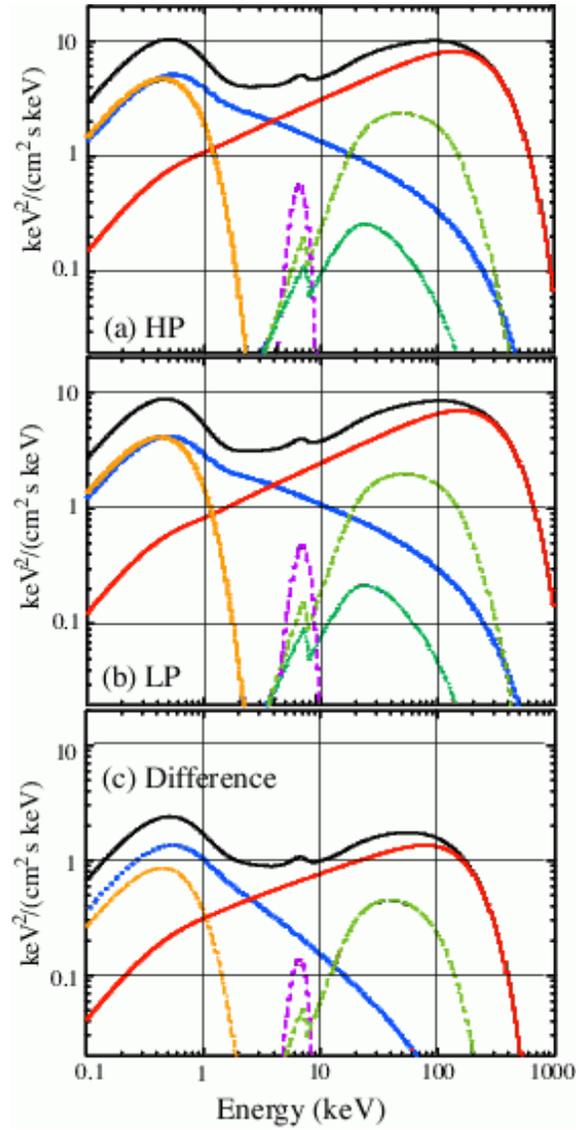}
\end{center}
 \caption{The same as figure~\ref{fig:bestfitmodels}a,
but obtained from  the HP (panel a) and LP (panel b) spectra.
Panel c is the fit to the HP minus LP difference spectra.}
\label{fig:HiLo_bestfitmodels}
\smallskip
\end{figure}

Following the approximate quantification
of the intensity-sorted HXD spectra (subsection~\ref{subsec:sortedspec}),
we fitted the full  (XIS2, PIN, and GSO) HP and LP spectra
with the  \texttt{wabs*(diskbb +compPS+compPS+gau)} model
which  has been found to be the best representation of 
the time-averaged spectrum (section~\ref{subsec:comppsfits}).
In the fitting, we fixed $N_{\rm H} = 6.6 \times 10^{21}$ cm$^{-2}$
as obtained with the average spectra.
In reference to the preliminary fittings conducted in the previous subsection,
we also fixed  all the reflection parameters to the values
obtained with the average spectra.
Then, as presented in figure~\ref{fig:HiLo_compPSfits},
the model has successfully reproduced both spectra.
The derived  parameters are listed in table~\ref{tbl:parameters},
in comparison with those from the averaged spectrum.
Thus, the LP and HP values of each free parameter 
(except a few rather unconstrained ones)
are found to generally bracket that from  the average spectrum.
The inferred best-fit models, in the $\nu F\nu$ form,
are given in figure~\ref{fig:HiLo_bestfitmodels}.

\begin{figure}[hbt]
\begin{center}
\FigureFile(8cm,){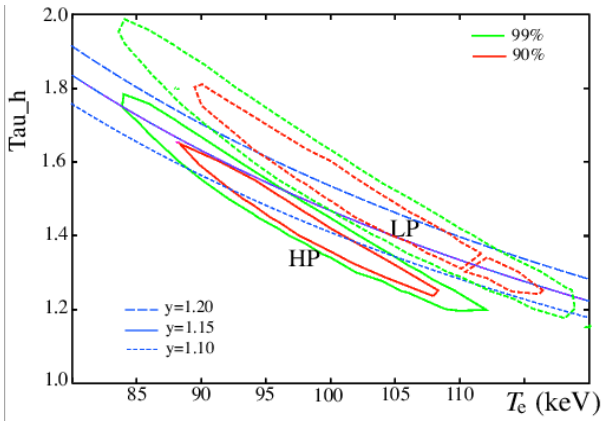}
 \end{center}
\caption{Confidence contours of the \texttt{compPSh} paramters in HP (solid lines)
and LP (dashed ones),
shown on the plane of $T_{\rm e}$ versus the optical depth $\tau_{\rm h}$.
Red and green contours represent 90\%  and 99\% confidence levels, respectively.
Blue lines indicate constant values of $y$.
\label{fig:HiLo_contours}}
\end{figure}

Table~\ref{tbl:parameters} and figure~\ref{fig:HiLo_bestfitmodels}
allow us to  derive several important inferences.
First, let us consider the observed spectral softening 
from LP to HP (figure~\ref{fig:HiLo_spec}b).
Although  this effect might suggest an increased relative contribution 
of \texttt{compPSs} in HP, this is not the case.
In fact, as the source varies,
the seed-disk radius of \texttt{compPSs} is kept at $3\pm 1$ times that of \texttt{compPSh}
(table~\ref{tbl:parameters}),
and their cross-over energy stays at $\sim 4$ keV (figure~\ref{fig:HiLo_bestfitmodels}).
In search for an alternative explanation, 
figure~\ref{fig:HiLo_contours} compares the \texttt{compPSh} parameters
between HP and LP  on the plane of $T_{\rm e}$ against the optical depth $\tau_{\rm h}$.
Thus, the HP and LP spectra become  distinct on this  plane,
with implication similar to what was derived in subsection~\ref{subsec:sortedspec};
the two phases cannot be represented by
a common set of  $T_{\rm e}$ and $\tau_{\rm h}$.
If, e.g.,  $T_{\rm e}$ is  the same, 
then $\tau_{\rm h}$ must  be lower in HP than in LP;
if  instead $\tau_{\rm h}$ is assumed to be the same,
$T_{\rm e}$ must be higher in LP.
The two phases are more clearly (though not completely) separated in $y$,
because it can be accurately specified by the observed spectral slopes.
We may, for simplicity,  express
that the hard continuum is ``less strongly Comptonized'' in HP than in LP. 
This is an important result enabled by the broad-band capability of Suzaku.
As to \texttt{compPSs}, 
we cannot find significant differences between LP and HP
beyond their relatively large errors.

Next, the source variation has been found to approximately conserve
the innermost disk temperature at $T_{\rm in} = 0.19$ keV (table~\ref{tbl:parameters}). 
In the present modeling, 
this quantity is assumed to be common to 
the directly visible \texttt{diskbb} component, 
and the seed photon sources for the two \texttt{compPS} components.
Therefore, it is determined mainly by local shapes 
of the spectral soft excess (figure~\ref{fig:HiLo_bestfitmodels}).
However, the broad-band capability is again very important,
because it allows us to accurately fix 
the Compton continua and the reflection hump,
and hence to unambiguously model  the soft excess.

As a third result from this analysis, 
we can attribute the LP to HP intensity increase
to an increase in  the spherically-averaged seed-photon luminosity,
which is given as 
\begin{equation}
L_{\rm seed} =4 \pi \sigma_0 T_{\rm in}^4 \left\{ {R_{\rm in-h}^{\rm seed}}^2 +{R_{\rm in-s}^{\rm seed}}^2 \right\}~;
\label{eq:Lseed}
\end{equation}
here, $\sigma_0$ is the Stefan-Boltzmann constant, 
while $R_{\rm in-h}^{\rm seed}$ and $R_{\rm in-s}^{\rm seed}$ respectively
represent the seed-disk radius of   \texttt{compPSh} and \texttt{compPSs}, 
as defined by equation (\ref{eq:seeddisk}).
When $L_{\rm seed}$ is estimated  
table~\ref{tbl:parameters} via equation~\ref{eq:Lseed},
its nominal error ranges overlap between HP and LP, 
due to large errors in $R_{\rm in}^{\rm seed}$.
However,  in reality, $R_{\rm in}^{\rm seed}$ is  tightly anti-correlated with $T_{\rm in}$ 
so as to  preserve the seed-photon flux (for \texttt{compPS} in particular).
Therefore, the combined error range of  $L_{\rm seed}$ becomes much smaller.
A detailed confidence-range examination yields
$L_{\rm seed} = (8.8^{+6.1}_{-0.8}) \times 10^{36}$ erg s$^{-1}$ in HP,
and  $(6.9^{+0.7}_{-1.8}) \times 10^{36}$ erg s$^{-1}$ in LP.

Finally,
the bolometric luminosity of the directly-visible \texttt{diskbb} component,
calculated after equation (\ref{eq:Lseed}) as
\begin{equation}
L_{\rm raw}=4 \pi \sigma_0 T_{\rm in}^4  {R_{\rm in}}^2 ~
\label{eq:Lraw}
\end{equation}
and presented in table~\ref{tbl:parameters},
is consistent with being the same between LP and HP,
within rather large  errors
which arise mainly via a strong coupling between \texttt{diskbb} and \texttt{compPSs}.
These error ranges  associated with $L_{\rm raw}$,
which also take into account the temperature vs. radius anti correlation,
are such that a significant decrease (by as much as a factor of $\sim 5$) 
in $L_{\rm raw}$ from LP to HP is allowed,
while its increases should not exceed $\sim 20\%$.
We therefore infer, as our 4th result, 
that the directly visible disk luminosity is  likely to be
less variable than the hard X-ray flux,
and there remains a possibility 
that it even anti-correlates with the hard X-ray flux.

\subsection{Difference spectra}
\label{subsec:diffspec}

To reconfirm and reinforce  the results  obtained in the preceding subsection,
we directly subtracted the raw (i.e., background-inclusive and dead-time uncorrected) 
LP spectra from those in  HP.
Since the source resides in either phase for a typical duration of several seconds
which is much shorter than the time scale of the NXB variation,
we expect the NXB to accurately cancel out by this subtraction,
making ``difference spectra"  free from the NXB modeling errors.
In fact,  over the highest energy of 530--600 keV
where the signal is considered negligible ($<1\%$ of the NXB),
the HP minus LP  count rate of HXD-GSO has been found
to be $(-0.6 \pm 1.0)\%$ of the time-averaged mean rate.
This also confirms 
that the potentially higher (by at most 2\%; subsection~\ref{subsec:sortedspec})
dead time in HP than in LP can be neglected.
The derived difference spectra are presented in figure~\ref{fig:HiLo_compPSfits}c.
As expected, they have several times smaller normalizations 
than the HP and LP spectra, and are generally softer.

As before,
we fitted the difference spectra using the progressively complex Comptonization models,
but fixing  $N_{\rm H}=6.6 \times 10^{22}$ cm$^{-2}$, $T_{\rm in}=0.19$ keV,
$\Omega/2\pi = 0.4$, $\xi=0$, and $R_{\rm in}^{\rm ref}=2500 R_{\rm g}$.
The systematic errors were not incorporated due to lower statistics.
Then, a single \texttt{compPS} continuum,
together with reflection and the \texttt{diskbb} component,
gave a poor fit with $\chi^2/\nu = 3.36$ for $\nu = 67$.
The fit goodness was improve to $\chi^2/\nu = 3.18$ ($\nu = 66$)
by activating the Gaussian component,
of which the center energy and the width are fixed at 
6.3 keV and  $\sigma=1.0$ keV, respectively.
Finally, by incorporating the second \texttt{compPS} component,
the fit has become fully acceptable ($\chi^2/\nu = 1.05$ for $\nu = 64$),
yielding the parameters given in table~\ref{tbl:parameters} 
(column before the last).
Thus, the model has again the same construction
as those for the time-averaged, HP, and LP spectra.
The best-fit model is presented in figure~\ref{fig:HiLo_compPSfits}c
in the convolved form,
and in figure \ref{fig:HiLo_bestfitmodels}c in the incident $\nu F \nu$ form.
Below, we examine the fit results.

In the difference spectra, the \texttt{compPSs} component has been found
to have again $\sim 3$ times larger seed-disk radius than  \texttt{compPSh}.
Since this ratio is close to those found in the average, HP, and LP spectra,
we reconfirm that the soft-photon inputs to the two Compton clouds
are varying roughly in proportion to each other.
In contrast, the value of $T_{\rm e} \sim 60$ keV is 
considerably lower than was found before.
One possible scenario from these results is  
that the Compton-cloud  temperature in fact has a distribution,
which becomes more weighted toward lower values as the source flares up,
so that the difference can be approximated by a single lower temperature.

In  subsection \ref{subsec:sortedspec}, 
we confirmed (though using the conventional model)
that the HP and LP spectra have  roughly the same $\Omega/2\pi$.
This is reinforced by the results obtained subsequently, 
namely the successful double-Compton modelings of 
the HP, LP, and the difference spectra, all with $\Omega/2\pi $ fixed at 0.4.
Therefore, the reflection component is thought to be fully 
catching up with the continuum variations on the 1-s time scale.
Also, the iron-line photon flux is varying  clearly on this time scale;
its variation  is consistent with being directly proportional to the continuum flux,
because all the four spectra of Cyg X-1 presented in table~\ref{tbl:parameters} have, 
within errors, the same Fe-K line equivalent width of $\sim 300$ eV .
These results are consistent with \citet{Revnivtsev1999}, 
who reported that  the reflection features (Fe-K line and smeared edge)
start  failing to follow continuum variations
above several Hz in frequency.

Although the fit to the difference spectrum incorporates the \texttt{diskbb} component,
the final model given in table~\ref{tbl:parameters} does not necessarily require it:
only an upper limit can be imposed on $R_{\rm in}$,
because the softest end of the difference XIS spectrum can be
accounted for by \texttt{compPSs} as well.
We in fact confirmed
that a negative normalization of \texttt{diskbb}
(i.e., a decrease of this component from LP to HP)
is allowed up to $\sim 40$\% (in absolute value) of those in  LP.
This is consistent with the results obtained in the previous subsection.

\subsection{Summary of spectral changes}
\label{subsec:sec4summary}

Through the analyses in the preceding three subsections,
the spectra changes associated with  the fast (1 to 200  s) intensity variations
have been quantified in the following manner.

\begin{enumerate}

\item
The intensity increase from LP to HP is explained
by an increase in the seed-photon luminosity, $L_{\rm seed}$.

\item
The spectral softening  from LP to  HP can be attributed
to weakening in the Comptonization of \texttt{compPSh}
(either small $\tau$, or lower $T_{\rm e}$, or both).

\item
Unlike  $L_{\rm seed}$,
the directly visible disk luminosity, $L_{\rm raw}$,
does not increase significantly in HP, 
and could even decrease.

\item
As the source varies,
the innermost temperature  of the cool disk is kept 
approximately constant at $T_{\rm in}=0.19$ keV.

\item
On these  time scales,
the Fe-K line and the reflection hump follow the continuum variations,
because $\Omega/2\pi$ and the Fe-K line EW 
are both consistent with being constant.

\end{enumerate}

\section{Discussion}
\label{sec:discussion}
%
\subsection{Summary of the Data Analysis
\label{subsec:summary_of_analysis}}

The Suzaku observation of Cyg X-1, 
conducted for an exposure of 17 ks 
at a 0.7--300 keV luminosity of $4.6 \times 10^{37}$ erg s$^{-1}$,
has provided one of the highest-quality LHS spectra of this prototypical BHB.
The XIS, HXD-PIN, and HXD-GSO spectra, 
altogether covering an extremely broad energy band of 0.7--400 keV,
have been explained simultaneously and consistently,
by invoking  two basic  constituents of the X-ray emission.
One is the hot Comptonizing corona with $T_{\rm e} \sim 100$ keV,
which manifests itself  in  the high-energy  spectral roll over
detected with a high significance by  HXD-GSO.
The other is the optically-thick cool disk with $T_{\rm in}\sim 0.2$ keV,
which produces the soft excess and the Fe-K line in the XIS band,
as well as the reflection hump in the HXD-PIN band
with $\Omega/2\pi \sim 0.4$.

The data are fully consistent with the cool disk 
supplying seed photons to the Compton cloud,
although other possibilities may not be necessarily excluded.
In the present modeling,
the soft photons from the cool disk are considered to reach us 
either directly as the soft excess (the \texttt{diskbb} component),
or through weaker Comptonization with 
$y \sim 0.3$ (\texttt{compPSs}),
or after experiencing  stronger Comptonization with $y \sim 1.15$ (\texttt{compPSh}).

Assembling together the two Compton components (\texttt{compPSh, compPSs})
and the three  cool-disk related ones (\texttt{diskbb}, reflection, and gaussian),
we have successfully reproduced the overall continuum shape,
including the complicated Fe-K line and soft-excess regions.
The Fe-K  line is inferred to be broadened 
only weakly to $\sim 1$ keV (in Gaussian $\sigma$), 
without any evidence for relativistically broadened wings
of which the detection is claimed from some BHBs  \citep{Miller_diskline}.
As seen in figure~\ref{fig:bestfitmodels},
the inferred incident spectrum shows too complicated a shape,
all over the energy range,
to be approximated  by a single power-law.

Analyzing in section~\ref{sec:ana_varspec} 
the characteristic flare-up behavior of Cyg X-1 on a time scale of $\sim 1$ s,
we found that the \texttt{wabs*(diskbb+compPS+compPS+gau)} modeling
applies also to the HP, LP, and the difference spectra.
The flare-up behavior can be understood primarily 
as an increase in the seed-photon  supply,
accompanied by a decreased Comptonization.
While the seed-disk luminosity thus increases in HP,
the luminosity of the directly visible \texttt{diskbb} 
does not increase significantly.
As the source varies,
$T_{\rm in}$ of the cool disk is kept  approximately constant.

\subsection{Comparison with the BeppoSAX results
\label{subsec:bepposax}}
Applying the double-\texttt{compPS} modeling to the BeppoSAX data of Cyg X-1 
in the LHS covering up to $\sim 200$ keV (whereas Suzaku reached 400 keV),
\citet{Frontera2001} derived $T_{\rm e} = 59 \pm 5$ keV,
$y=0.89 \pm 0.01$, and $\Omega/2\pi = 0.25 \pm 0.04$
for \texttt{compPSh} (their  Compton component 1), 
and $T_{\rm e} = 42 \pm 19$ keV and $y=0.15 \pm 0.01$ 
for \texttt{compPSs}  (their \texttt{copmPS} component 2).
Except for some differences in the model parameters,
the Suzaku and  BeppoSAX  results are hence consistent
in that the  broad-band spectra can be reproduced 
by a pair of Comptonized components.
The two results grossly agree on the  soft-excess parameters as well,
although \citet{Frontera2001} used a blackbody 
to model the soft excess and the seed-photon source.

The BeppoSAX observation was conducted
at  a 0.5--200 keV unabsorbed flux of $4.2 \times 10^{-8}$ erg cm$^{-2}$ s$^{-1}$,
or a luminosity of $3.1 \times 10^{37}$ erg s$^{-1}$ 
(in the same band at a distance of 3.2 kpc).
The corresponding values from the present Suzaku observation are higher,  
$5.9 \times 10^{-8}$ erg cm$^{-2}$ s$^{-2}$ ($4.4 \times 10^{37}$ erg s$^{-1}$).
Then, we would expect Suzaku to observe a lower $T_{\rm e}$,
based on the  general understanding
that $T_{\rm e}$ decreases toward higher luminosities
due to enhanced Compton cooling (e.g., \cite{Esin1998,Yamaoka2005,ZG2004}).
Nevertheless, we in fact measured a 1.6 times 
{\em higher} $T_{\rm e}$ than BeppoSAX.
One possible explanation to this behavior is
that Cyg X-1 may not have fully returned to the LHS at the time of the BeppoSAX observation,
which was conducted soon after the source transition from the high/soft state.

Prior to the present work, the double-Compton model was applied 
successfully  to wide-band LHS spectra of Cyg X-1 
(\cite{G1997,Frontera2001,Ibragimov2005}; 
though some explicitly invoking multiple values of $T_{\rm e}$)  
and GRO~J1655$-$40 (Paper~I, subsection~\ref{subsec:1655_cyg}).
Therefore, we  regard this model as a promising description of the LHS spectra of BHBs.
As discussed in Paper~I,
the mixture of two Compton optical depths
(plus the direct \texttt{diskbb} component)
allows either a ``spatial'' or a ``time-domain'' interpretation.
In the former case, 
some fraction of the cool  disk is considered directly visible,
while the remaining  part is covered by a hot electron cloud
that has thinner and thicker portions.
The latter (time domain) alternative assumes 
that the Compton cloud is rapidly varying among three typical conditions, 
represented by $\tau=0$, $\tau \sim 0.4$, and $\tau \sim 1.5$.
The actual condition may be a combination of these two simplified cases.
Of course, as pointed out previously (e.g., Paper~I), 
the double-Compton modeling may be an approximation 
to more complex conditions involving
more than two  Compton optical depths,
or even a continuous distribution in $\tau$
as suggested by some models (e.g., \cite{Liu2002}).
With the present data, 
it is virtually impossible to distnguish which is more likely,
a continuous distributin in $\tau$ or just two optical depths.

\subsection{Comparison with GRO~J1655$-$40
\label{subsec:1655_cyg}}
A comparison between  Cyg X-1 (with $i \sim 45^\circ$)
and  GRO~J1655$-$40 (with $i \sim 70^\circ$) would help 
elucidating  the source geometry  (section~\ref{sec:intro}). 
After a brief attempt made in Paper~1
using the BeppoSAX results on Cyg X-1 \citep{Frontera2001},
here we perform a more accurate comparison 
between the two objects using the Suzaku data on both.
For this purpose, figure~\ref{fig:bestfitmodels}b reproduces
the best-fit  \texttt{wabs*(diskbb+compPS+compPS+gau)} model for GRO~J1655$-$40, 
deduced in Paper~I using  the Suzaku data. 
On that occasion (2005 September 22 and 23), 
GRO~J1655$-$40  was in the LHS
with a 0.7--300 keV  luminosity of $4.9 \times 10^{36}$ erg s$^{-1}$ (Paper~I),
which  is an order of magnutude lower than that of  Cyg X-1 
in the present observation (both assuming isotropic emission).
However, the difference becomes a factor of 3--5
when the luminosity is normalized to the Eddington value,
because the BH in Cyg X-1  has a mass of $12-20~M_\odot$,
while that in  GRO~J1655$-$40 has $\sim 6.5~M_\odot$  (Paper~I).

\begin{figure}[thb]
\begin{center}
\FigureFile(8cm,){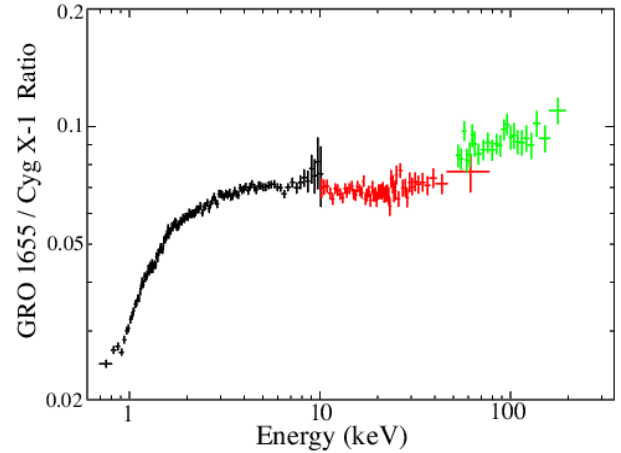}
 \end{center}
\caption{Background-subtracted  XIS (black),
 HXD-PIN (red), and HXD-GSO (green) spectra of GRO J1655-40 (Paper~I),
 divided by those of Cyg X-1 obtained in the present study.}
\label{fig:1655_cyg_specratio}
\end{figure}

Although the two  models in figure~\ref{fig:bestfitmodels} 
generally look  alike, they differ in details.
To grasp such  differences in a model-independent manner,
we divided the XI2, HXD-PIN, and HXD-GSO spectra 
of  GRO~J1655$-$40 by the corresponding spectra of Cyg X-1,
and obtained the results shown in figure~\ref{fig:1655_cyg_specratio}. 
Below, we attempt to identify basic features in this  spectral ratio,
and  to interpret them in terms of the spectral decomposition
in figure~\ref{fig:bestfitmodels} and table~\ref{tbl:parameters}.
Like in figure~\ref{fig:HiLo_spec}b, 
the behavior of the ratio at the highest XIS energy range should be ignored, 
where the GRO~J1655$-$40 data may suffer from photon pile-up effects (Paper~I).

The  spectral ratio in figure~\ref{fig:1655_cyg_specratio} is 
roughly constant over  $\sim 3$ to $\sim 30$ keV,
but decreases significantly below $\sim 2$ keV.
This effect is too large to be explained by their difference in absorption,
$N_{\rm H} = 6.6 \times 10^{21}$ cm$^{-2}$ in Cyg X-1
and $7.4\times 10^{21}$ cm$^{-2}$ in GRO~J1655$-$40.
Instead, it can be attributed to the stronger
soft excess of Cyg X-1 (figure~\ref{fig:bestfitmodels}),
as the \texttt{diskbb} component of Cyg X-1 is 
an order of magnitude more prominent  than that of GRO~J1655$-$40
when normalized to the overall continuum
(while $T_{\rm in}$ is nearly the same).
This in turn  may be ascribed primarily to their difference in the $\cos i$ factor,
$\cos(45^\circ)/\cos(70^\circ)=2.1$,
because we expect the direct disk emission to scale as $\propto \cos i$,
while the Comptonized emission to be approximately isotropic.
Strictly speaking,  this  estimate needs some caution.
In subsection~\ref{subsec:HiLo_compPSfits},
we introduced two quantities, $L_{\rm raw}$ and $L_{\rm seed}$.
Because $R_{\rm in}$  used to calculate $L_{\rm raw}$
has already been corrected for the $\cos i$ factor 
while $R_{\rm in}^{\rm seed}$ is not
(subsection~\ref{subsec:comppsfits}, table~\ref{tbl:parameters}),
we expect the ratio $L_{\rm raw}/L_{\rm seed}$ 
to be approximately inclination independent.
Nevertheless, the ratio was measured to be $\sim 0.44$ in Cyg X-1,
while $< 0.19$ in GRO~J1655$-$40 (table~\ref{tbl:parameters}).
That is, the soft excess in GRO~J1655$-$40 is 
even weaker than would be predicted by the assumed $i=70^\circ$.
This is discussed later in subsection~\ref{subsec:variation}.

In figure~\ref{fig:1655_cyg_specratio}, 
we observe a small dip at $\sim 6.5$ keV.
Evidently, this feature arises 
because the Cyg X-1 spectrum exhibits a stronger 
Fe-K line (with an equivalent width of  $\sim 300$ eV)
than that of GRO~J1655$-$40 ($\sim 80$ eV).
This property is again ascribed, at least qualitatively, 
to the inclination effects:
the cool flat disk will intercept a fraction of the Comptonized photons,
and produce the fluorescent Fe-K photons,
of which  the visibility must be proportional to $ \cos i$.
Quantitatively,  however, the difference in the equivalent width between 
the two sources amounts to a factor of $\sim 4$,
which exceeds the factor $\sim 2$ difference  expected from $i=70^\circ$.
This is also discussed  in subsection~\ref{subsec:variation}.

Over the 30--200 keV range,
the ratio in figure~\ref{fig:1655_cyg_specratio} increases 
with energy the  $E$ roughly as  $\propto E^{0.3}$:
GRO~J1655$-$40 has a harder continuum.
This is consistent with  the PL approximation to their  hard X-ray spectra,
which gave $\Gamma=1.75$ for  Cyg X-1 (section~\ref{subsec:convfits})
and $\Gamma=1.39$ for GRO~J1655$-$40 (Paper~I).
In terms of the Comptonization scenario,
this slope difference can be attributed to the higher $y$-parameter
of \texttt{compPSh} in GRO~J1655$-$40  ($1.31^{+0.11}_{-0.08}$)
than in Cyg X-1 ($1.15^{+0.05}_{-0.03}$).
This in turn can be explained rather securely 
by the higher $T_{\rm e}$ of GRO~J1655$-$40,
unlike the case of the HP vs. LP comparison of Cyg X-1
in which the origin of the change in $y$ remained ambiguous.
As discussed in Paper~I and subsection~\ref{subsec:bepposax},
the higher $T_{\rm e}$ of GRO~J1655$-$40 may be due to its lower luminosity. 

In the PIN energy range, 
the ratio of figure~\ref{fig:1655_cyg_specratio} exhibits a  slightly concave curvature.
Because the continuta themselves in this energy range
should be  simple power-laws dominated by \texttt{compPSh},
this effect must be caused by  differences in the reflection features.
In fact, figure~\ref{fig:bestfitmodels} reveals two effects.
One is that the reflection hump of Cyg X-1, relative to the Compton continuum,
is  inferred to be stronger than that in GRO~J1655$-$40.
This is  again considered an   inclination effect,
because the reflected component is expected to be proportional to $\cos i$
while the primary continuum is thought to be  rather isotropic.
Actually, the values of $\Omega/2\pi$ in table~\ref{tbl:parameters},
which is calculated  {\it after correction} for $\cos i$,
becomes comparable between the two objects.
The other effect is that the reflection hump in Cyg X-1,
 modeled with  $i = 45^\circ$, emerges  at 30--50 keV,
whereas that of  GRO~J1655$-$40, modeled with $i = 70^\circ$, 
is peaked at higher energies.
This is because a larger $i$ makes
the reflection more forward directed, 
and hence the reflected photons suffer less Compton degradation \citep{MZ1995}.
These two effects in combination are considered
to produce the slightly concave-shaped spectral ratio in the PIN band.
Incidentally, the comparable values of $\Omega/2\pi$
between the two objects disfavors the idea
that the hot Comptonizing corona may have a semi-relativistic
outflow velocity away from the disk plane  \citep{Malzac2001},
because in that case we would observe a smaller value of $\Omega/2\pi$
from Cyg X-1 due to the enhanced continuum.

To summarize, 
the spectrum of Cyg X-1 and GRO~J1655$-$40 are both dominated by the Comptonized continua,
but Cyg X-1 exhibits more prominently the flat-disk-related spectral features 
presumably due to its lower inclination.
Another inclination effect manifests itself in
the  subtle differences in the 10--70 keV continuum shape, 
presumably due to inclination-dependent changes in the 
strength and peak energy of the reflection hump.
These results are consistent with the view
that the Comptonizing hot cloud is geometrically thick,
and hence its emission  depends much less on the inclination
than  the flat-disk related components.

Analyzing  broad-band BeppoSAX spectra of XTE~J1118+480,
\citet{Frontera2001b} discovered that this transient BHB has 
a very weak soft excess and a weak reflection hump.
Like the present discussion on GRO~J1655$-$40,
these properties of XTE~J1118+480 may  also be ascribed to the inclination effects,
since this object  is also reported  to  have a rather high inclination
as  $i=(68 \pm 2)^\circ$
\citep{Gelino2006}.

\subsection{The disk parameters in the low/hard state}
\label{subsec:diskpara}

In the high/soft state of Cyg X-1,
a \texttt{diskbb} fit to the ASCA data showed
the optically-thick disk to have $r_{\rm in}= 71$~km \citep{Dotani1997},
which translate via equation~(\ref{eq:rin2Rin})
to $R_{\rm in}=90$~km for $i=30^\circ$ \citep{Makishima2000}
or $R_{\rm in}= 100$~km for $i=45^\circ$.
The disk is securely considered to reach the last stable orbit
\citep{Dotani1997,Makishima2000}.
In the LHS, a similar cool disk is  considered to be present as well,
but its conditions become significantly more difficult to estimate,
because its emission is overwhelmed and strongly modified by the Compton process.
As a result, there is considerable controversy over the innermost disk radius in the LHS.

Theoretically, the Advection Dominanted Accretion Flows  \citep{ADAF95} 
and other related hot flow models,
meant to explain the LHS,  all assume 
that the cool disk is truncated
at some radius  larger than the last stable orbit. 
(One important difference is that the original ADAF model
assumes synchrotron photons as the soft seed photons 
to be Compton up-scatterd into hard X-rays,
whereas the present scenario assumes the cool disk 
to supply  the seed soft photons.)
A gradual inward shift in this truncation radius will
give rise to the correlated spectral softening,
increased reflection, and increased variation power frequencies,
as observed  (e.g., \cite{Gilfanov1999,ZLS1999,Z2004,ZG2004,Done2007}).
When the disk reaches the last stable orbit,
a  transition to the high/soft state will take place.
However, this view has been challenged by some LHS observations 
which suggest the disk to extend down to the last stable orbit 
\citep{Miller2006,MHM2006,Rykoff2007}. 
Here we use the broad-band Suzaku data to infer the disk parameters in the LHS.

In  subsection~\ref{subsec:convfits},
the conventional  \texttt{diskbb+cutoffpl} fit to the present LHS data
tentatively gave $r_{\rm in} ~\sim 13$ km,
and hence $R_{\rm in}= 18$~km for $i=45^\circ$.
This apparently implies that the cool disk extends 
down to a much smaller  radius than in the high/soft state.
However,  as pointed out before \citep{Ebisawa1996,DiSalvo2001,Ibragimov2005}
and revealed by the present data (figure~\ref{fig:bestfitmodels}),
the LHS spectra of Cyg X-1 and other BHBs are significantly more complex 
than is described by a single power-law.
In fact, our final solution invoking double \texttt{compPS}
has given an order of magnitude larger value, 
$R_{\rm in} = 250$ km (table~\ref{tbl:parameters}),
which is consistent with the disk inner radius receding 
by a factor 2.5 compared to the high/soft state.
The large difference in $R_{\rm in}$  between the two modelings
for the same data can be explained in the following manner.
In our final model, the flux in the 1--2 keV range is mostly explained
by  \texttt{compPSs} (figure~\ref{fig:bestfitmodels}a),
with \texttt{diskbb} carrying softer residuals.
If, instead, a single straight continuum is employed,
the \texttt{diskbb} component becomes unusually hot 
trying to explain the 1--2 keV excess
\citep{DiSalvo2001,Ibragimov2005},
and hence the inner disk radius becomes apparently too small.
This gives a lesson
that disentangling the disk parameters,
where the disk is not the dominant component in the spectrum, 
is subject to large uncertainties in the underlying continuum shape. 
Evidently, the case of GRO~1655$-$40 is subject to larger difficulties
because of the weaker soft excess.

The  Comptonization scenario  involves another important effect 
which artificially reduces the disk radius:
the disk photons are ``lost''  from the \texttt{diskbb} component
into the dominant Comptonized continua.
In estimating the true disk radius,
to be denoted $R_{\rm in}^{\rm tot}$,
we must  take this effect into account,
knowing that Comptonization conserves the photon number
\citep{KME2001,Aya+Max2004,Aya+Chris2004}. 
Practically, $R_{\rm in}^{\rm tot}$ can be derived by quadratically summing 
$R_{\rm in}$ of \texttt{diskbb},
$R_{\rm in}^{\rm seed}$ of \texttt{compPSh}, 
and $R_{\rm in}^{\rm seed}$ of \texttt{compPSs} as
\begin{equation}
\left(R_{\rm in}^{\rm tot} \right)^2 = R_{\rm in}^2 
                   + \left(R_{\rm in-h}^{\rm seed}\right)^2
                   +\left(R_{\rm in-s}^{\rm seed}\right)^2~~.
\label{eq:Rtot}
\end{equation}
Then, equations~ (\ref{eq:Lseed}),  (\ref{eq:Lraw}), and (\ref{eq:Rtot})
readily yield the total luminosity $L_{\rm disk}$ of the cool disk  as 
\begin{equation}
  L_{\rm disk} = L_{\rm raw}+L_{\rm seed}
                 =4\pi \left({R_{\rm in}^{\rm tot}} \right)^2 \sigma_0 T_{\rm in}^4~.
\label{eq:Ldisk}
\end{equation}
From the values  in table~\ref{tbl:parameters}  for the time-averaged spectrum,
we  obtain $R_{\rm in}^{\rm tot} \sim 330$ km,
which is equivalent to  $\sim 15R_{\rm g}$ for $M_{\rm BH}=15~M_\odot$.
The implications are
that roughly half the disk area is directly visible,
and the disk inner radius is  factor 3 larger than in the high/soft state.

Although the estimated value of $R_{\rm in}^{\rm tot}$ may still be 
relatively small compared with the original idea of disk truncation,
it  can be subject to yet additional uncertainties
which  take part in 
when transforming the raw \texttt{diskbb} normalization  into a physical radius. 
The stress-free inner boundary condition is probably appropriate 
for a thin disk extending down to the last stable orbit. 
In contrast, a truncated disk may have continuously rising stress to its inner radius. 
Similarly, the environment of the disk is different between the two states:
in the LHS, the disk is likely to be more strongly heated 
by irradiation and conduction from the hot cloud.
Therefore, the color hardening factor in the LHS  may well be significantly larger 
than the value of 1.7 employed in equation~(\ref{eq:rin2Rin});
then, the true effective temperature would be lower,
and hence $R_{\rm in}$ would become still larger.

Three independent pieces of evidence further
argue against the small disk radius in the LHS.
One is  the relatively small values of reflection, $\Omega/2\pi \sim 0.4$,
which requires the cool seed disk not to intrude too deeply into the hot corona.
To reconcile the small values of $\Omega/2\pi$ with  a configuration
wherein  the seed disk is nearly completely surrounded by the corona,
we would have to invoke, e.g., mild relativistic coronal outflows (e.g., \cite{Malzac2001}),
but we already argued against it (subsection~\ref{subsec:1655_cyg}).
Second, the intrinsic width of the Fe-K line,
when interpreted as relativistic broadening in the disk near the BH,
can constrain the innermost locations of the cool disk.
For this purpose, we re-fitted the average spectra with the final model,
but with the Gaussian replaced by so-called \texttt{diskline} model
\citep{Fabian89}.
When fixing the inclination  at $45^\circ$,
and assuming the line emissivity to  scale as $r^{-3}$,
the fit constrained the rest-frame line center energy as $6.3_{-0.1}^{+0.2}$  keV,
and the innermost disk radius for the Fe-K line production
as $13_{-7}^{+6} \; R_{\rm g}$  which is fully consistent 
with the value of $R_{\rm in}^{\rm tot}$ derived above.
Finally,  as quoted in subsection~\ref{subsec:diffspec},
\citet{Revnivtsev1999} studied fast variations of Cyg X-1 in the Fe-K line energy region,
and argued  the cool reprocessor to be located farther than 
$\sim 100 R_{\rm g}$ for $M_{\rm BH}=15~M_\odot$.
Although the implied radius is much larger than considered here,
this should be taken as another evidence supporting  the disk truncation.
(Incidentally, the present results on the fast variability in 
the Fe-K lines and the Compton humps constrain
 their production sites to be closer than $ 1.3 \times 10^4~R_{\rm g}$.)

Based on these examinations,
we estimate that the cool-disk radius of Cyg X-1 
during the present LHS observation 
is $R_{\rm in}^{\rm tot} \sim 15R_{\rm g} \sim 330$ km nominally,
but is subject to rather large uncertainties,
and is likely to be considerably larger than this.

All the considerations presented here will also apply to previous LHS studies,
where the small inner disk radius derived from simple continuum fits 
was used to challenge the truncated disk model
\citep{Miller2006,MHM2006,Rykoff2007}. 
In short, the small disk radii claimed for some BHBs 
in the LHS would need a careful revision.

\subsection{A possible interpretation of the fast X-ray variation}
\label{subsec:variation}

The remaining issue is how to interpret our results on the time variation,
summarized in subsection~\ref{subsec:sec4summary}
and subsection~\ref{subsec:summary_of_analysis}.
In a standard accretion disk 
which we believe is approximately correct in the present case,
the disk parameters are mutually related as 
$T_{\rm in } \propto (R_{\rm in}^{\rm tot})^{-3/4} \dot{M}^{1/4}$ 
(e.g., \cite{Makishima2000}), 
where $\dot{M}$ is the mass accretion rate.
Since  $T_{\rm in}$ has been confirmed to change little as  Cyg X-1 varies,
the variation must also keep  $R_{\rm in}^{\rm tot}$ and $\dot{M}$ 
 approximately constant.
(An alternative case, namely, 
positively correlated changes in $R_{\rm in}^{\rm tot}$ and $\dot{M}$,  
would be unlikely,
because an increase in $\dot {M}$ would enhance radiative cooling,
and hence {\em reduce} $R_{\rm in}^{\rm tot}$;
\cite{ZLS1999,ZG2004,Done2007}.)
As  a result, the total disk luminosity $L_{\rm disk}$,
defined by equation~(\ref{eq:Ldisk}), should also be kept rather constant.
However, $L_{\rm seed}$  has been found to  increase 
from LP to HP (subsection~\ref{subsec:HiLo_compPSfits}).
Therefore,  $L_{\rm raw}$ must decrease in HP by 20--30\%.
As  confirmed in subsections \ref{subsec:HiLo_compPSfits} and \ref{subsec:diffspec},
the data permit (though do not require) such a decrease 
in $L_{\rm raw}$ from LP to  HP.

Based on the above arguments, we presume that Cyg X-1 
gets brighter as a larger fraction of the disk photons are
intercepted by the Comptonizing corona,
while the underlying cool disk itself is kept essentially unchanged.
This will cause an increase in $L_{\rm seed}$,
and the associated  decrease in $L_{\rm raw}$.
Although this inference does not necessarily specify
how the corona is actually changing, 
we may consider one possible scenario below.

Let us, for example, assume 
that the Compton cloud has a rather spherical shape, or a thick torus geometry, 
with its outer radius  located at $R_{cc} \sim 100~R_{\rm g} \sim 2 \times10^3$ km
where the free-fall time is $\sim 0.1$ s.
(The actual in-fall time scale of the corona may well be longer,
depending on the viscosity.)
Based on our discussion made in subsection~\ref{subsec:diskpara},
the cool disk is considered to  partially protrude into the corona,
but truncated at, say, $R_{\rm in}^{\rm tot}  \sim 50 ~R_{\rm g}$.
The Comptonization will then take place both in the overlapping region
(between $R_{\rm in}^{\rm tot}$ and $R_{\rm cc}$), 
and inside $R_{\rm in}^{\rm tot}$, of the corona.
The corona is assumed to be rather turbulent with strong density inhomogeneities,
so that we need two optical depths,
$\tau_{\rm s} \sim 0.4$ and $\tau_{\rm s} \sim 1.5$,
to express its effects.
Furthermore, we expect $T_{\rm e} $ to increase as the corona falls toward the BH,
for the following two reasons.
One is because,
compared to the corona inside $R_{\rm in}$,
that in the overlapping region must be more strongly cooled by the disk photons,
and less efficiently heated by protons due to lower densities.
The other is because,  at about $R_{cc}$, 
the disk surface layer must evaporate into the corona,
so it will take a fraction of a second (calculated at $R_{\rm cc}$; \cite{ei_coupling})
for the evaporated electrons to be heated by protons via Coulomb interaction.
Then, the  observed $T_{\rm e}\sim 100$ keV should be 
considered as a radial average over the corona.

Under the configuration as assumed above, 
we further speculate that, for some unspecified reasons,
the corona has a number of  ``holes''
(at least in the overlapping region between the disk and corona),
through which the disk is directly visible (subsection~\ref{subsec:bepposax}).
With these, the  spectral composition of Cyg X-1 
(and of GRO~J1655$-$40) can be  explained at least semi-quantitatively.
Furthermore, this scenario can explain
why the directly visible disk and the Fe-K lines of GRO~J1655$-$40
are weaker than those of Cyg X-1
beyond what is expected by the inclination difference,
because the holes would act as a vertical collimator to direct 
these disk-related emission components toward pole-on directions.
The disk reflection would also be affected,
but thier intrinsically broad shape would remain rather intact.

In order to explain the fast variability,
we may assume that the opening fraction of the coronal holes vary with time,
due to  some fluctuations in the disk evaporation rate.
As the hole area diminishes by an increased evaporation,
we expect $L_{\rm seed}$ to increase and $L_{\rm raw}$ to decrease,
as observed in the changes from LP to HP.
Since the incremented portion of the corona 
must initially have a lower electron temperature
due to the finite ion vs. electron coupling time, 
the HP spectrum is expected to be less Comptonized than in LP,
again as observed.
Electrons in that portion of the corona will be gradually heated up,
on time scales of a fraction of a second,
as they fall inwards across $R_{\rm in}^{\rm tot}$.
This will  explain, at least qualitatively, the XIS to PIN phase lag 
which  would be too large to be explained 
by  time-of-flight delays of photons in the Compton process.

\section{Summary}
\label{sec:summary}

Through the detailed analysis of the Suzaku data of Cyg X-1,
we  have obtained the following firm results.
\begin{enumerate}
\item
The  0.7--400 keV Suzaku spectra are reproduced 
by a sum of a cool disk emission,
two Compton continua both with reflection ($\Omega/2\pi \sim 0.4$), 
and a mildly broad  Fe-K line. 

\item 
The  two Compton continua, with $\tau \sim 0.4$ and $\sim 1.5$,
can be described by a common temperature of  $T_{\rm e} \sim 100$ keV,
although the data do not exclude them having different temperatures. 

\item
The disk is characterized by $T_{\rm in} \sim 0.2$ keV
and  $R_{\rm in} \gtrsim 330$ km ( $\gtrsim 15~R_{\rm g}$).
Roughly half the  disk emission gets  Comptonized,
while the rest forms the spectral soft excess. 

\item
When Cyg X-1 brightens up on a time scale of 1--200 s,
a larger fraction of disk photons becomes Comptonized,
although the disk parameters remain unchanged.
The spectrum softens,
as a result of a decrease in either $T_{\rm e}$ or $\tau$.

\item
A comparison between Cyg X-1 (with $i \sim 45^\circ$) 
and GRO~J1655$-$40 (with $i \sim 70^\circ$) supports the view
that the disk has a flat geometry and the Compton cloud is more spherical.
\end{enumerate}

Combining these results with some theoretical considerations,
we propose the following speculative scenario on the LHS of Cyg X-1
(and possibly of other BHBs).
That is, the Componizing corona, with a roughly spherical (or thick torus) geometry,
extends up to, say, $\sim 100 R_{\rm g}$,
and the cool disk protrudes  half way into it.
The corona is highly inhomogeneous (as required by the double Compton $\tau$),
and has openings (as required by the soft excess) 
to allow a direct view of part of the disk. 
The hard X-ray intensity increases 
when the opening area of the corona decreases
and hence a larger number of disk photons become Comptonized.

\bigskip

{\it Acknowledgement} \
The authors would like to express their hearty thanks to
all the members of the Suzaku Science Working Group,
for their help in the spacecraft operation,
instrumental calibration, and data processing.
This work was partially supported by the Grant-in-Aid for Scientific Research 
on Priority Areas (Grant No. 14079101).

{}

\onecolumn
\begin{table}
\caption{Simultaneous fits to the  XIS2, HXD-PIN, and HXD-GSO spectra of Cyg X-1 
using two thermal Comptonization components.$^*$
} 
\label{tbl:parameters}
\begin{center}
\begin{tabular}{lcccccc}
\hline \hline
Component & Parameter &   Average   & High Phase  &  Low Phase & Difference & GRO~J1655$-$40 $^\dagger$ \\
\hline
wabs        &  $ N_{\rm H}$ (10$^{21}$ cm$^{-2}$) &  $6.6^{+0.8}_{-0.3}$     & (6.6 fix)      & (6.6 fix)   & (6.6 fix)    & $7.4^{+0.3}_{-0.1}$\\
\hline
diskbb      &  $T_{\rm in}$ (keV)   &  $0.19^{+0.01}_{-0.02}$ &$0.19 \pm 0.01$ &$0.18 \pm 0.01$& (0.19 fix) & $0.18 \pm 0.01$ \\
  & $R_{\rm in}$ (km) $^\ddagger$ & $250^{+210}_{-60}$    & $250^{+30}_{-150}$ & $ 270^{+50}_{-40}$ &$<110$& $< 73$ \\
                & $L_{\rm raw} ~^\S$         & $10^{+13}_{-3}$       & $11^{+1}_{-9}$    & $10 \pm 1$       &$<2$ & $< 1$   \\
\hline
compPSh &  $T_{\rm e}$ (keV) &  $100 \pm 5$                 &  $95^{+15}_{-5} $        &  $105^{+10}_{-15}$  & $60^{+10}_{-15}$ & $140 \pm 10$ \\
    &  optical depth          &  $ 1.49^{+0.07}_{-0.09}$ &$ 1.5^{+0.2}_{-0.3}$   & $ 1.4^{+0.4}_{-0.1}$  & $2.0 ^{+0.5}_{-0.4}$& $ 1.14 \pm 0.08$\\
   & $y$-parameter        & $1.15^{+0.05}_{-0.03}$ & $1.12^{+0.03}_{-0.09}$ & $1.17^{+0.11}_{-0.03}$ & $0.89 \pm0.3$ & $1.31^{+0.11}_{-0.08}$\\
& $R_{\rm in}^{\rm seed}$ (km)$^\|$ & $75 \pm 10$    & $80^{+15}_{-10}$   & $75^{+10}_{-15}$  &$45 \pm 5$ &  $26^{+5}_{-4}$ \\
\hline
compPSs$^\#$ 
    &  optical depth  &  $0.38^{+0.04}_{-0.05}$ &$0.38^{+0.07}_{-0.22}$ & $0.32^{+0.21}_{-0.05}$  & $0.4 \pm 0.2$ & $0.25^{+0.02}_{-0.03}$\\ 
    & $y$-parameter   & $0.29^{+0.04}_{-0.03}$   &$0.28^{+0.03}_{-0.14}$   
                                          & $0.27^{+0.11}_{-0.03}$  &$0.16 ^{+0.09}_{-0.04}$& $0.28^{+0.07}_{-0.05}$\\
& $R_{\rm in}^{\rm seed}$ (km)$^\|$ & $200^{+90}_{-30}$    
                                         & $210^{+80}_{-10} $  & $210^{+40}_{-20}$ &$130 ^{+15}_{-30}$ &$64^{+7}_{-10}$\\
\hline
reflection$~^\P$ &  $\Omega/2\pi$ &  $0.4^{+0.2}_{-0.3}$  &(0.4 fix)    & (0.4 fix)        & (0.4 fix) &  $ 0.5 \pm 0.1$ \\
                &  $\xi$ (erg cm s$^{-1}$)   &  $< 30$    & (0 fix)         & (0 fix)       & (0 fix)    & $400 \pm 100$\\        
                & $R_{\rm in}^{\rm ref} (R_{\rm g})$
                                                   &  $2500^{\L}$   & (2500 fix)     &  (2500 fix)       & (2500 fix) & $> 200$\\
\hline
Gaussian & centroid (keV)         &  $6.3 \pm 0.1$    & $6.3^{+0.1}_{-0.2}$   & $6.3^{+0.1}_{-0.2}$ & (6.3 fix) & $6.3^{+0.1}_{-0.2}$ \\
                & eq. width (eV)        &  $290 \pm 90$    &  $290 \pm 50$    & $330^{+60}_{-50}$    & $400 \pm 200$ & $100^{+10}_{-20}$ \\
                & sigma (keV)            &  $1.0^{+0.1}_{-0.2}$   & $0.9^{+0.2}_{-0.1}$    & $1.1 \pm 0.2$   & (1.0 fix)& $0.7 \pm 0.1$ \\ 
\hline
fit goodness& $\chi^2_\nu~(\nu)$ & 1.13 (349)     &   1.08 (228) &  1.17 (193)    &  1.01 (64) & 1.03 (1484) \\
\hline
figure      &                   & figure~\ref{fig:comppsfits}a,d 
                                       & figure~\ref{fig:HiLo_compPSfits}a
                                         & figure~\ref{fig:HiLo_compPSfits}b 
                                            & figure~\ref{fig:HiLo_compPSfits}c &             Paper I\\
\hline
\end{tabular}
\end{center}
\medskip
\begin{itemize}
\item[$^*$]  Cyg X-1 and GRO~J1655$-$40 are assumed to be at distances of 2.5 kpc and 3.2 kpc, respectively.
The fitting model is \texttt{wabs*(diskbb+compPS+compPS+gauss)}.
\item[$^\dagger$]  The results on GRO~J1655$-$40 refer to  Paper~I.
\item[$^\ddagger$]  The innermost disk radius derived via equation~(\ref{eq:rin2Rin}),
  assuming $i=45^\circ$ for Cyg X-1 and $i= 70^\circ$ for GRO~J1655$-$40. 
\item[$^\S$] Bolometric luminosity  in $10^{36}$ erg s$^{-1}$, 
     obtained as $L_{\rm raw}=4\pi  R_{\rm in}^2 \sigma_0 T_{\rm in}^4$ 
     with $\sigma_0$ the Stefan-Boltzmann constant.
\item[$^\|$]  Square root of the \texttt{compPS} normalization, 
  converted to the inner radius of the seed photon disk via  equation~(\ref{eq:seeddisk}).
  Note this is not divided by the $\sqrt{\cos i}$ factor.
\item[$^\#$] Assumed to have the same $T_{\rm e}$ as \texttt{compPSh}. 
\item[$\P$] The parameters are assumed to be common to the two \texttt{compPS} components.
 The reflector inclination is assumed in the same way as in $^\ddagger$.
\item[$\L$] Not constrained in the fitting.
\end{itemize}
\end{table}

\begin{table}
\caption{Cutoff power-law fits to the  HXD (PIN and GSO) spectra in HP and LP.$^*$} 
\label{tbl:HiLo_convfits}
\begin{center}
\begin{tabular}{lcccc}
\hline \hline
Component& Parameter      &   Free   & Common $\Gamma$  &  Common $E_{\rm cut}$         \\
\hline
cutoff PL     & $\Gamma$ (HP)    & $1.51 \pm 0.02$ & $1.51 \pm 0.01$  &$1.52 \pm 0.01$ \\
                   & $\Gamma$ (LP)     & $1.50 \pm 0.02$ &   ---   $^\dagger$  &$1.49 \pm 0.01$ \\
            \cline{2-5}
	           & $E_{\rm cut}$ (HP) & $212 \pm 12$& $208^{+5}_{-8}$& $220 \pm 10$\\
	           & $E_{\rm cut}$ (LP)  & $234 \pm 16$ & $240\pm 12 $& ---$^\dagger$          \\
            \cline{2-5}
                    & Norm. ratio$^\ddagger$  &   1.31                     & 1.27               & 1.39 \\
\hline
fit goodness & $\chi_\nu~(\nu)$      & 0.93 (168)             & 0.93 (169)           & 0.94 (169) \\
 \hline \hline
\end{tabular}
\end{center}
\begin{itemize}
\item[$^*$] The HP and LP spectra were fitted simultaneously with the \texttt{cutoffpl+pexrav} model.
The reflection strength was fixed at $\Omega/2\pi=0.16$
\item[$^\dagger$]  Constrained to be the same between the HP and LP spectra.
\item[$^\ddagger$]  The ratio of \texttt{cutoffpl} normalization (at 1 keV) between HP and LP.
\end{itemize}
\end{table}

\end{document}